\documentclass[twocolumn]{aastex63}

\newcommand{\zdec}{z_\mathrm{dec}}
\newcommand{\amax}{a_\mathrm{max}}
\newcommand{\amin}{a_\mathrm{min}}
\newcommand{\aset}{a_\mathrm{mix}} 
\newcommand{\hg}{H_\mathrm{g}}
\newcommand{\hd}{H_\mathrm{d}}
\newcommand{\hfloor}{H_\mathrm{floor}}
\newcommand{\sigd}{\Sigma_\mathrm{d}}
\newcommand{\au}{\mathrm{au}}

\usepackage{bm}
\usepackage[T1]{fontenc}
\usepackage{tabularx}
\usepackage{threeparttable}
\usepackage{graphicx}
\usepackage{comment}

\graphicspath{{./figs/}}
 
\shorttitle{JWST observations of the HH 30 disk}
\shortauthors{Tazaki et al.}

\begin{document}

\title{JWST Imaging of Edge-on Protoplanetary Disks. IV. Mid-infrared Dust Scattering in the HH 30 disk}

\correspondingauthor{Ryo Tazaki}
\email{ryo.tazaki1205@gmail.com}

\author[0000-0003-1451-6836]{Ryo Tazaki}
\affiliation{Univ. Grenoble Alpes, CNRS, IPAG, F-38000 Grenoble, France}
\affiliation{Department of Earth Science and Astronomy, The University of Tokyo, Tokyo 153-8902, Japan}

\author[0000-0002-1637-7393]{Fran\c{c}ois M\'enard}
\affiliation{Univ. Grenoble Alpes, CNRS, IPAG, F-38000 Grenoble, France}

\author[0000-0002-5092-6464]{Gaspard Duch\^ene}
\affiliation{Univ. Grenoble Alpes, CNRS, IPAG, F-38000 Grenoble, France}
\affiliation{Astronomy Department, University of California, Berkeley, CA 94720, USA}

\author[0000-0002-8962-448X]{Marion Villenave}
\affiliation{Universit\'a degli Studi di Milano, Dipartimento di Fisica, via Celoria 16, 20133 Milano, Italy}

\author[0000-0003-3133-3580]{\'Alvaro Ribas}
\affiliation{Institute of Astronomy, University of Cambridge, Madingley Road, Cambridge CB3 0HA, UK}

\author[0000-0002-2805-7338]{Karl R. Stapelfeldt}
\affiliation{Jet Propulsion Laboratory, California Institute of Technology, 4800 Oak Grove Drive, Pasadena, CA 91109, USA}

\author[0000-0002-3191-8151]{Marshall D. Perrin}
\affiliation{Space Telescope Science Institute, Baltimore, MD 21218, USA}

\author[0000-0001-5907-5179]{Christophe Pinte}
\affiliation{School of Physics and Astronomy, Monash University, Clayton Vic 3800, Australia}
\affiliation{Univ. Grenoble Alpes, CNRS, IPAG, F-38000 Grenoble, France}

\author[0000-0002-9977-8255]{Schuyler G. Wolff}
\affiliation{Department of Astronomy and Steward Observatory, University of Arizona, Tucson, AZ 85721, USA}

\author[0000-0001-5334-5107]{Deborah L. Padgett}
\affiliation{Jet Propulsion Laboratory, California Institute of Technology, 4800 Oak Grove Drive, Pasadena, CA 91109, USA}

\author[0000-0003-3583-6652]{Jie Ma}
\affiliation{Univ. Grenoble Alpes, CNRS, IPAG, F-38000 Grenoble, France}

\author[0009-0001-8094-4735]{Laurine Martinien}
\affiliation{Univ. Grenoble Alpes, CNRS, IPAG, F-38000 Grenoble, France}

\author[0009-0003-4758-4285]{Maxime Roumesy}
\affiliation{Univ. Grenoble Alpes, CNRS, IPAG, F-38000 Grenoble, France}

\begin{abstract}
We present near- and mid-infrared (IR) broadband imaging 
observations of the edge-on protoplanetary disk around HH 30 
with the James Webb Space Telescope/Near Infrared Camera (NIRCam) and the Mid-Infrared Instrument (MIRI). We combine these observations with archival optical/near-IR scattered light images obtained with the Hubble Space Telescope (HST) and a millimeter-wavelength dust continuum image obtained with the Atacama Large Millimeter/submillimeter Array (ALMA) with the highest spatial resolution ever obtained for this target. 
Our multiwavelength images clearly reveal the vertical and radial segregation of micron-sized and sub-mm-sized grains in the disk. 
In the near- and mid-IR, the images capture not only bi-reflection nebulae separated by a dark lane but also diverse dynamical processes occurring in the HH 30 disk, such as spiral- and tail-like structures, a conical outflow, and a collimated jet. In contrast, the ALMA image reveals a flat dust disk in the disk midplane. By performing radiative transfer simulations, we show that grains of about 3 \micron~in radius or larger are fully vertically mixed to explain the observed mid-IR scattered light flux and its morphology, whereas millimeter-sized grains are settled into a layer with a scale height of $\gtrsim1$ au at $100$ au from the central star.
We also find a tension in the disk inclination angle inferred from optical/near-IR and mm observations with the latter being closer to an exactly edge-on. Finally, we report the first detection of the proper motion of an emission knot associated with the mid-IR collimated jet detected by combining two epochs of our MIRI 12.8-\micron~observations. 
\end{abstract}

\section{Introduction}

Planets form in a protoplanetary disk. Growth of micron-sized dust grains followed by vertical settling and radial drift of grains is a crucial step in the formation of millimeter-sized or larger grains \citep[pebbles;][]{Pinilla21}, which is one of the key prerequisites for planetesimal formation \citep[see][for a recent view]{Birnstiel24}. Pebbles in the outer region of some Class I/II disks are highly settled \citep{Pinte16, Villenave22}, whereas there also exist disks showing a moderate/minimal degree of dust settling in Class 0/I disks \citep{Villenave23, DanielLin23, Villenave24, Guerra24} or even in a Class II disk \citep{Doi21, Doi23}. It is still unclear from the observational point of view how and when the vertical settling of dust grains occurs in disks.

Disks seen from nearly edge-on view, so-called edge-on protoplanetary disks, are a unique laboratory to study the vertical dust settling and radial drift of dust grains in disks from observations. If an edge-on disk is observed at a sufficiently high spatial resolution, we may distinguish between light coming from different heights of the disk, providing more direct clues on the vertical dust distribution as well as their radial disk sizes. 
In particular, multiwavelength observations from optical/near-infrared (IR) to millimeter wavelengths allow us to study the change in the vertical/radial distribution of grains as a function of grain size, as we can probe different grain sizes at different wavelengths \citep{Duchene24}.

As such, Herbig Haro 30 (HH 30) is the prototypical edge-on protoplanetary disk and
is one of the most intriguing targets because of a wealth of dynamical processes happening in it, such as vertical stratification of small and large dust grains \citep{Burrows96, Villenave20}, variable disk-scattered light \citep{Burrows96, Stapelfeldt99, Watson07, Duran2009}, collimated jets \citep{Mundt83, Burrows96, Anglada07, Estalella12, Pascucci24}, and low-velocity molecular outflow  \citep{Pety06, Louvet18, Lopez24, Pascucci24}. 
HH 30 belongs to the L1551 region \citep{Lynds1962} in the Taurus star-forming cloud, located at a distance of $146.4\pm0.5$ pc \citep{Galli19}. 
Photometric properties of HH 30 suggest that it is an accreting T-Tauri star in a Class II phase with a spectral type of M0$\pm2$ \citep{White04}. 
The mass of the central star has been estimated to be $\sim0.45 M_\sun$ based on the Keplerian rotation of CO emissions \citep{Pety06, Lopez24}.
It is worth noting that HH 30 has been suspected of being a binary system to account for the wiggling of the optical jet \citep{Anglada07, Estalella12}.

The HH 30 disk has been imaged with a high spatial resolution at two limiting wavelengths.
One is at optical and near-IR wavelengths with the HST, where we probe sub-micrometer-sized grains at the disk surface \citep{Burrows96,Stapelfeldt99, Watson04}. At these wavelengths, the disk is observed as bi-reflection nebulae separated by a dark lane. The other one is at millimeter (mm) wavelengths with Plateau de Bure Interferometer (PdBI) and ALMA \citep{Pety06, Guilloteau08, Louvet18,Villenave20}. In this case, thermal emission of pebbles in the disk midplane is observed, revealing a flat and compact disk morphology \citep{Louvet18, Villenave20}, although the vertical thickness was not resolved in the previous PdBI and ALMA images \citep{Pety06, Louvet18}.
These two observational wavelengths differ by a factor of a thousand, and the morphology of the disk at wavelengths in between remains unknown.

Thanks to its high spatial resolution and exquisite sensitivity, JWST offers the opportunity to study these missing intermediate wavelengths for the first time.
While we were at the final stage of our paper preparation, \citet{Pascucci24} reported JWST/NIRSpec/IFU observations of four edge-on disks including HH 30, and found a nested structure of outflow emission lines extending northeast of the disk.
Here, we present broadband imaging observations at near- to mid-IR of HH 30 with JWST/NIRCam and the MIRI. We also report the dust continuum observations with ALMA Band 6 with the highest spatial resolution ever obtained for this target. By combining these observations with HST observations, we study the multiwavelength appearance of the HH 30 disk, allowing us to constrain the dynamical evolution of dust grains in the disk. 
This paper is organized as follows. 
We summarize our observations and data reduction in Section \ref{sec:obs} and the main observational results are presented in Section \ref{sec:res}. In Section \ref{sec:model}, we describe our radiative transfer model, and the results are presented in Section \ref{sec:modelres}.
Discussion and Conclusion are given in Sections \ref{sec:discussion} and \ref{sec:conclusion}, respectively.

\section{Observations and Data Reduction} \label{sec:obs}

\subsection{JWST} \label{sec:reducjwst}

\begin{table}[t]
\caption{Observing Log of the Dataset Used in This Study}
\label{tab:obslog}
\centering
\begin{tabular}{lllll} 
\hline\hline          
Telescope & Instrument & Filter/Band & $t_\mathrm{int}$  & UT Date\\ 
\hline
 HST & NICMOS & F110W & 256s& 1997/09/29  \\
 \ldots & \ldots & F160W & 256s& 1997/09/29  \\
 \ldots & \ldots & F204M & 448s& 1997/09/29  \\
 \ldots & WFPC2 & F439W & 2400s& 1998/12/01  \\
 \ldots & \ldots & F555W & 900s& 1998/12/01  \\
 \ldots & \ldots & F675W & 4800s& 1998/12/01  \\
 \ldots & \ldots & F814W & 1800s& 1998/12/01  \\
 \ldots & \dots & F555W & 2100s& 2008/11/10 \\
 \ldots & \ldots & F675W & 4200s& 2008/11/10  \\
 \ldots & \ldots & F814W & 2100s& 2008/11/10  \\
 JWST & MIRI & F770W &  1118s& 2023/01/23  \\
 \ldots & \ldots & F1280W & 1132s& 2023/01/23  \\
 \ldots & \ldots & F2100W & 1159s& 2023/01/23  \\
 \ldots & NIRCam & F200W & 429s& 2023/09/07  \\
 \ldots & \ldots & F444W & 429s& 2023/09/07  \\
 \ldots & MIRI & F770W  & 1118s& 2023/09/07  \\
 \ldots & \ldots & F1280W & 1132s& 2023/09/07  \\
 \ldots & \ldots & F2100W & 1159s& 2023/09/07  \\
 ALMA & & Band 6 & 1h10min & 2013/07/19-21 \\
 \ldots & & Band 6 & 1h30min & 2017/10/07 \\
\hline
\end{tabular}
\tablecomments{
The fourth column represents the total on-source integration time.}
\end{table}   

We observed HH 30 with JWST as part of GO program 2562 (PI F. M\'enard, K. Stapelfeldt), using NIRCam and MIRI in two consecutive visits.
The observations were first carried out on 2023 January 23. The observation was incomplete due to a guide star acquisition failure, and we only obtained MIRI images with the F770W, F1280W, and F2100W filters. The observations were repeated on 2023 September 7th, providing both NIRCam (F200W and F444W) and MIRI (F770W, F1280W, and F2100W) images. See Table \ref{tab:obslog} for more details, including the total integration time. 

We obtained the level-2 reduced data products from the Mikulski Archive for Space Telescopes (MAST). The retrieved files were then processed using the JWST pipeline version 1.12.0. We ran the level-3 processing in order to subtract the sky background. To do this, we set \texttt{subtract=True} along with the default sky match method \texttt{`global+match'} in the \texttt{skymatch} step in the pipeline. 
In the NIRCam/F200W image, the $1/f$ noise from the SIDECAR ASICs (detector readout electronics) is noticeable, which appears as striping artifacts in the image. To mitigate this stripe artifacts, we used the \texttt{image1overf} package\footnote{\url{https://github.com/chriswillott/jwst}}. We applied it to the corresponding \texttt{\_cal.fits} files for each detector quadrant and frame, creating stripe-subtracted level-2 files, which were then used as inputs of the level-3 pipeline.

All the pipeline-reduced images still contain spatially inhomogeneous background emission, which is either due to star-forming clouds or instrumental artifacts. Similar to what has been done in our series of papers \citep{Villenave24}, we perform 2D background subtraction using the \texttt{Background2D} function in the \texttt{photutils} python package.
For both NIRCam and MIRI images taken on 2023 September 7th, one of the diffraction spikes produced by a nearby bright source (XZ Tau), located northeast of HH 30, overlaps the HH-30 disk. We performed subtraction of the diffraction spike from the images by creating a model of the spike by fitting the intensity profile along it and using it to subtract the spike contamination. The detailed procedures are described in Appendix \ref{sec:spikesub}.

After performing the background and diffraction spike subtractions, we aligned the images by using a point source commonly visible among the images to compensate for a possible intra-instrument misalignment. 
We fitted the point source with a 2D Gaussian function to derive its central location and shifted the images to match the location of the same source seen in a reference image. 
For the September data, we used the F200W image as the reference and shifted the F444W, F770W, and F1280W images to match the source location in the reference. 
The resulting amount of shifts for F444W, F770W, F1280W are $0\farcs07$, $0\farcs08$, $0\farcs08$, respectively.
No alignment was applied to the F2100W image, as we found no point sources in the field of view.
Because of unknown astrometric accuracy, we do not use the September F2100W image when discussing the position of the emission.
For the January MIRI images, we used the aligned September-F444W image as a reference and shifted the January F770W, F1280W, and F2100W images by $0\farcs22$, $0\farcs23$, $0\farcs18$ to match the source location in the reference image, respectively. 
The values are similar to the expected pointing accuracy of JWST without science target acquisition of $\sim0\farcs1$ \footnote{\url{https://jwst-docs.stsci.edu/jwst-observatory-characteristics/jwst-pointing-performance\#gsc.tab=0}}.
Finally, we rotated each image by $31.6^\circ$ \citep{Anglada07} to bring the jet axis oriented upward. 

\subsection{HST data}

We obtained archival datasets of HH 30 via enhanced data products from the Hubble  Legacy  Archive  (HLA) from the MAST portal (Table \ref{tab:obslog}) for the following programs: Program 7228 executed on 1997/09/29 (PI: E. Young), Program 6754 executed on 1998/12/01 (PI: K. Stapelfeldt), Program 11867 executed on 2008/11/10 (PI: J. Trauger). 
For the images of the F555W and F814W filters obtained in Program 6754 and all filters obtained in Program 11867, HH 30 was placed on the PC1 chip of the WFPC2 camera, providing a pixel scale of 0\farcs0455/pixel, while other WFPC2 cameras WF2, WF3, and WF4 have a coarse pixel scale of 0\farcs0997/pixel.
The datasets from Programs 6754 and 7228 have been reported previously \citep{Cotera01, Watson07}, while those from Program 11867 are reported here for the first time except for the F675W image, which has been reported in \citet{Ai24}.
The images presented here are the \texttt{drz} product with the 2D background subtraction (Section \ref{sec:reducjwst}), as done for JWST images. We manually aligned HST images because of the lack of obvious background stars that can be used to improve absolute astrometry. Potentially inaccurate centering of the HST image is not crucial in our later discussion. 

\subsection{ALMA data}

We downloaded two sets of ALMA observations of HH~30 at 1.3 mm \citep[Band 6; ][]{Ediss04}, one using a compact antenna configuration (2013.1.01175.S, PI.: Dougados) and a second one with a more extended configuration (2017.1.01701.S, PI.: Villenave). The compact data set was observed three times betweeen 19 and 21 July 2013 with antenna baselines ranging from 15 to 1574 m. One of the executions was observed under considerably worse weather conditions, and we chose not to not use it in this work. The correlator setup included three narrowband spectral windows for different $^{12}$CO(2-1) isotopologues as well as two broader spectral windows for continuum (bandwidth of 1.875 GHz, 128 channels each), and we used the latter for our analysis. In the case of the extended configuration, two observations were obtained on 7 October 2017 with baselines from 40 m to 15.4 km. The correlator setup included three broad spectral windows (bandwidth of 1.875 GHz, 128 channels each), as well as a narrower one centered on the $^{12}$CO(2-1) line (bandwidth of 0.938 GHz, 1920 channels). To maximize the signal to noise of the extended configuration, we used the four spectral windows after flagging the channels containing $^{12}$CO(2-1) emission.

The standard pipeline calibration was applied to the compact data set using CASA version 4.3.1. The two executions were imaged together using {\tt tclean}, and a single round of phase-only selfcalibration was applied to each of them by averaging in time for the duration of each scan (i.e., using {\tt solint=inf} in {\tt gaincal}). The phase centers were then set to the center of the disk as determined from visual inspection of the image. Similarly, the observations taken with the extended antenna configuration were processed using the pipeline and CASA version 5.1.1. We then imaged and selfcalibrated them with a single round of phase-only calibration, this time generating a single solution for all spectral windows and the whole duration of the observations to increase the signal-to-noise of the solutions (i.e., using {\tt solint=inf} and combining {\tt spw} and {\tt scan} in {\tt gaincal}). As in the previous case, the phase center was then set to the visual center of the disk. The combination of the compact and extended configurations was performed by first assigning the same coordinates to the phase center of each observations, and then applying a round of phase-only selfcalibration to the whole data set to ensure proper alignment. Finally, we imaged the joint data set with {\tt tclean} using {\tt robust} values of 0.5 (synthesized beam of 0.029$\times$0.02\arcsec, PA=6 deg, and rms=9.1 $\mu$Jy/beam) and 1 (synthesized beam of 0.046$\times$0.023\arcsec, PA=20 deg, and rms=8.4 $\mu$Jy/beam) as trade-offs between angular resolution and sensitivity.

We rotated the ALMA image with the same angle as the other images. By fitting the jet seen in the NIRCam 2-\micron~image, we found that the jet axis and the axis of symmetry of the disk in the ALMA image are offset by $\sim0\farcs17$, which is likely due to a combination of the limited JWST's pointing accuracy and the self-calibration process of the ALMA image.
Since there is no background source in the field of view of ALMA that can be used to align the images, we applied a shift in the rotated image so that the symmetry axis of the ALMA image matches the jet axis seen in the 2-\micron~image. 

\section{Results} \label{sec:res}
\begin{figure}[tbp]
\centering
\includegraphics[width=\linewidth]{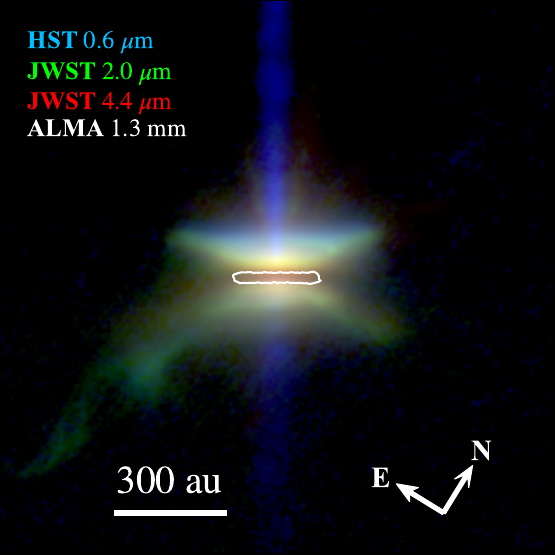}
\caption{An RGB color composite image of HST 0.6~\micron~(blue), JWST 2 \micron~ (green), 4.4 \micron~(red), and ALMA 1.3 mm (white contours) images. The HST and JWST images are shown with a logarithmic stretch with a common scale, each normalized to the peak value. The ALMA contour corresponds to the $5\sigma$ values. The field of view of the image is 10\arcsec$\times$10\arcsec. The image is rotated such that the jet (a position angle of $31.6^\circ$) is pointing upward. }
\label{fig:rgb}
\end{figure}

Figure \ref{fig:rgb} shows an RGB color composite image of HST and JWST images of the edge-on disk around HH 30 overlaid with the contours of the dust continuum emission observed with ALMA. The image demonstrates the complementarity of these three observatories to trace different regions of the disk, such as disk reflection nebulae with HST and JWST, a collimated jet with HST, and thermal emission from large grains in the disk midplane with ALMA.
All JWST images obtained in our program are shown in Appendix \ref{sec:jwstall}. 

\subsection{Disk scattered light with HST and JWST} \label{sec:ref}

\subsubsection{Overview}

\begin{figure}[tbp]
\centering
\includegraphics[width=\linewidth]{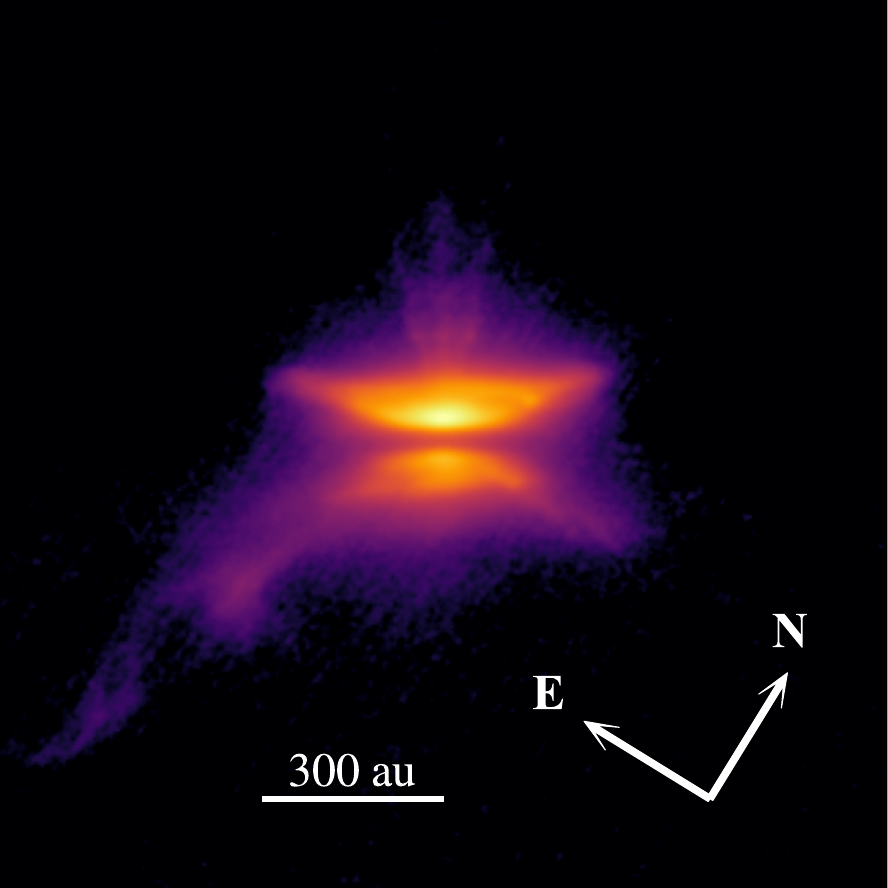}
\includegraphics[width=\linewidth]{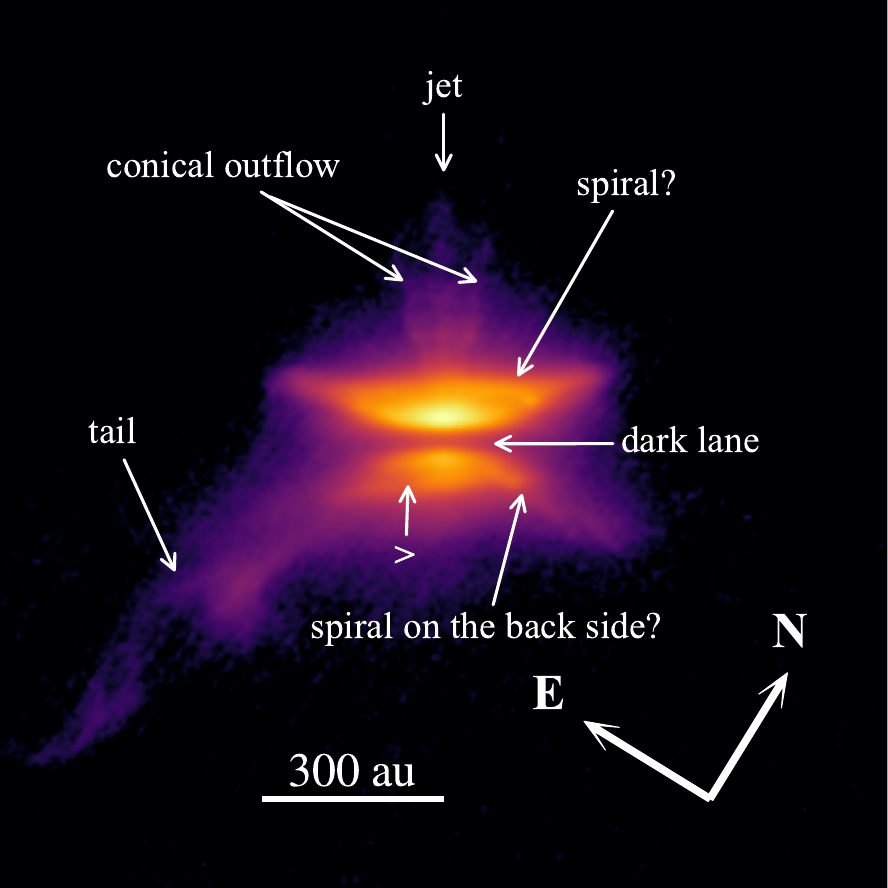}
\caption{
The JWST/NIRCam 2-\micron~image with a logarithmic stretch without (top) and with annotations to highlight major features seen in the image (bottom). The field of view of the image is 10\arcsec$\times$10\arcsec. }
\label{fig:nircam}
\end{figure}

\begin{figure}[tbp]
\centering
\includegraphics[width=\linewidth]{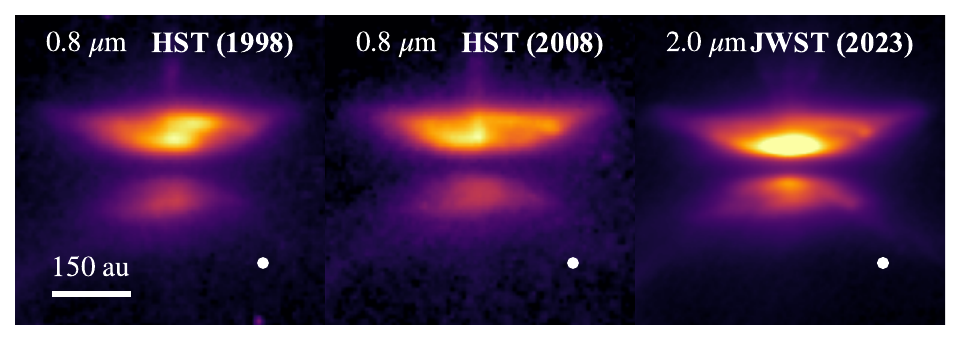}
\caption{Spiral-like structure seen at three epochs: 1998 and 2008 (HST/WFPC2-PC1/F814W) and 2023 (JWST/NIRCam/F200W). All images are shown with a square root stretch from zero to the peak value except for the JWST image, which we applied a further stretch to stand the spiral out.}
\label{fig:spiral}
\end{figure}

\begin{figure*}[tbp]
\centering
\includegraphics[width=\linewidth]{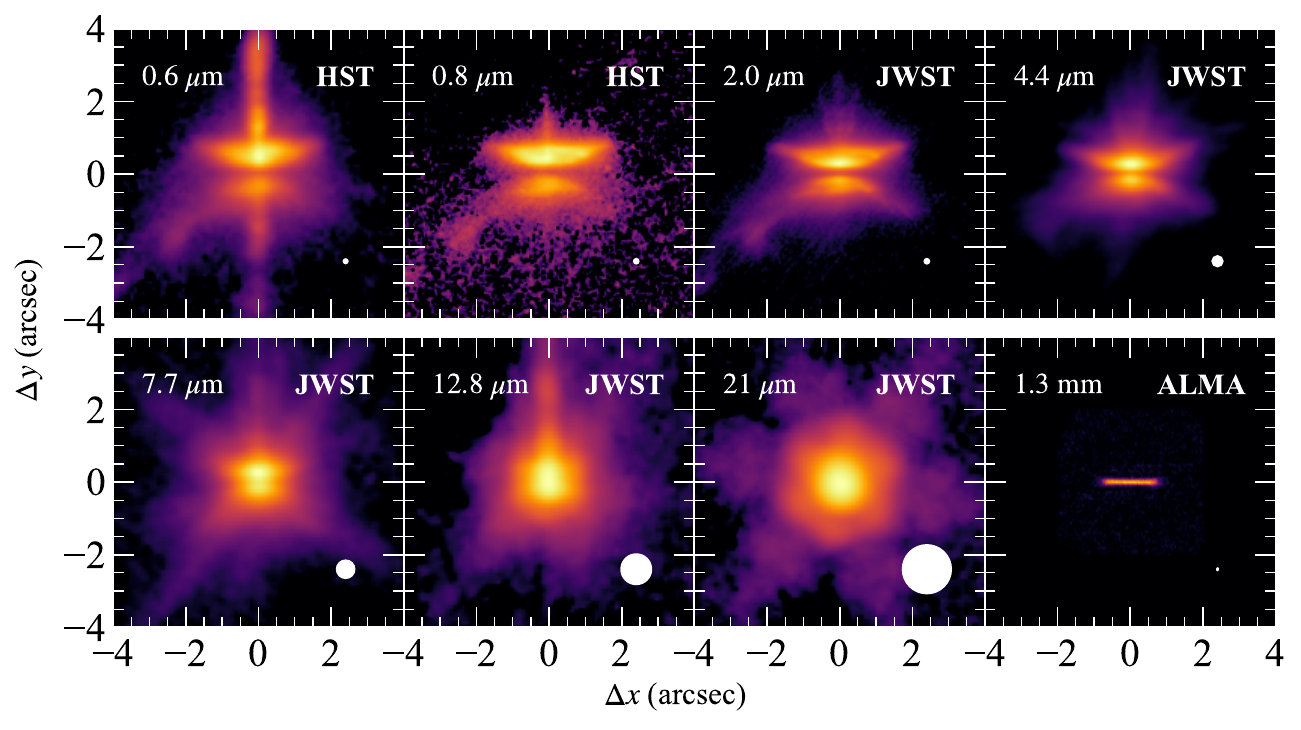}
\caption{Image gallery of the HST and JWST images with a logarithmic stretch and the ALMA Band 6 image with a linear stretch. All of them are shown with the same angular scale and the corresponding beam is shown in the bottom right of each panel. The diagonal feature seen in the 7.7-\micron~image is a detector artifact, known as the cruciform artifact \citep{Gaspar21}.}
\label{fig:hstjwstalma}
\end{figure*}

Figure \ref{fig:nircam} shows the JWST/NIRCam/2-\micron~image with a logarithmic stretch to highlight fainter features. The image clearly reveals bi-reflection nebulae separated by the dark lane.
Among features seen in the image, a conical feature is newly identified in broadband observations, although it is also seen in recently published NIRSpec/IFU images \citep{Pascucci24}. 
The spiral-like feature was seen only vaguely in previously published HST images \citep{Burrows96, Cotera01, Watson07}, whereas our JWST and HST (taken in 2008) images reveal it even more clearly (Figure \ref{fig:spiral}).
There exists a structure on the bottom surface that is possibly connected to the spiral seen on the top surface. A crab-claw-shaped feature (marked with ">") can also be seen in the bottom surface. The image also captures the jet and tail, although those were already detected in the early HST images \citep{Burrows96}.

Figure \ref{fig:hstjwstalma} shows the multiwavelength images of the HH 30 disk.
The disk reflection nebula seen in the JWST images has a nebular separation (or a vertical thickness of the dark lane) narrower than those seen in the HST images. 
The reflection nebulosity is visible up to 7.7 \micron~or perhaps 12.8 \micron. 
At 21~\micron, the disk is barely resolved because of a comparable size of the point spread function (PSF). 
The diagonal feature seen in the MIRI 7.7~\micron~image is the cruciform artifact \citep{Gaspar21} and thereby should not be considered as real emission (see Appendix \ref{sec:jwstall}). 
Disk substructures, such as the tail and spiral, are not clearly visible for longer wavelengths, but the jet and conical structure are still visible in the 4.4~\micron~and 7.7~\micron~images. 
The jet is particularly bright in the 12.8~\micron~filter, which includes an atomic line emission of [NeII], and the counter jet is not visible unlike in the 0.6-\micron~optical image.
To account for the absence of the counter-jet emission in the 12.8-\micron~filter, it needs to be fainter by a factor of at least $\sim30$ than the main jet at an altitude of ~2\farcs5.
The conical structure is also not visible in the 12.8-\micron~filter likely due to the lower resolution.
The mid-IR jet contains an emission knot, which appears to move away from the star between the two MIRI observing epochs from which we can measure the proper motion (Section \ref{sec:jet}).

Table \ref{tab:HH30_jwstphoto} summarizes the JWST photometry of HH 30. 
\citet{Cotera01} reported a flux of 1.9 mJy at 2 \micron~(with the HST/F204M filter), and our F200W flux is 26\% higher than this value. The difference is likely due to time variability, as \citet{Watson07} found a similar amount of difference ($\sim24\%$) at 1.6~\micron~(HST/F160W) between 1997/Sep/29 and 2004/Oct/7 with the former being fainter.
The MIRI fluxes between the two epochs are slightly higher in January than in September by 8\% at 21 \micron \footnote{The version of the JWST pipeline we used does not correct for the count-rate loss of the MIRI imager over time. By using the fitting curve of the count-loss rate reported on March 19, 2024 (\url{https://www.stsci.edu/contents/news/jwst/2024/an-improved-correction-for-the-miri-imager-long-wavelength-count-rate-loss-is-now-available}), 
we found the difference in the F2100W-filter flux due to the change of the count-loss rate is $\sim2\%$ between our two epochs of MIRI observations. The impact is therefore small compared to the absolute photometric calibration error ($\sim3\%$).}.
It is known from optical observations that the photometric flux of HH 30 varies with a timescale of 7.58 days with an amplitude as large as 77\% of the median flux during the K2 observations \citep{Cody22}.
Despite the variable nature of HH 30, the mid-IR fluxes of HH 30 are stable to within 10\% between the two epochs. This is true not only for fluxes but also for the morphology of the reflection nebulae, such as lateral or vertical intensity profile. 
The reflected light in the optical/near-IR is illumination by the star, which has spots and flares creating lots of short-term variability \citep{Stapelfeldt99}, while the reflected light in the mid-IR is illumination from the infrared excess of the inner disk which is not expected to vary nearly as much. It is worth noting however that the similarity of the two MIRI fluxes could merely be coincidence, and more epochs are useful to establish a mid-IR variability if present. 

\begin{table}[t]
    \centering
    \begin{threeparttable}
    \caption{JWST Photometry of HH 30}
    \begin{tabular}{ccc}
        \hline
        Date & Wavelength ($\mu$m) & Flux (mJy) \\
        \hline
        2023/01/23 & 7.7 & $2.4\pm0.07$\\
                   & 12.8 & $3.0\pm0.09$\\
                   & 21.0 & $10.6\pm0.3$\\
        \hline
        2023/09/07 & 2.0 & $2.4\pm0.05$\\
                   & 4.4 & $3.2\pm0.1$\\
                   & 7.7 & $2.4\pm0.07$ \\
                   & 12.8 & $2.9\pm0.09$\\
                   & 21.0 & $9.8\pm0.3$ \\
        \hline
    \end{tabular}
    \begin{tablenotes}[flushleft]
    \item \raggedright {\textbf{Notes.}  The flux is measured within the field of view of $12\arcsec\times12\arcsec$. The error bars are based on the absolute flux calibration errors, which are $\sim2\%$, $\sim4\%$, and $\sim3\%$ for F200W, F444W, and the MIRI filters, respectively, based on the JWST user documentation.}
    \label{tab:HH30_jwstphoto}
    \end{tablenotes}
    \end{threeparttable}
\end{table}

\subsubsection{Radial and vertical extents of the disk} \label{sec:extmor}

To characterize the observed nebular morphology, we measure the separation between the reflection nebulae, the flux ratio between the two surfaces, and their radial sizes as a function of wavelength.

To this end, we fitted the top and bottom spines of the surfaces using the method described in \citet{Duchene24}, which is briefly summarized below. For each image, we created vertical cuts at various separations from the disk's symmetry axis. To improve the signal-to-noise ratio, each vertical cut was averaged over a $\sim0.4$\arcsec~binning window. The vertical cut shows two peaks corresponding to two surfaces, each of which was fitted with a polynomial function except for the MIRI images. In the MIRI images, we used two Gaussian functions to fit the whole vertical profile because the intensity peaks of the two surfaces are blended, making it difficult to fit each surface separately. Once the positions of the spines were found from polynomial/two-component Gaussian fitting, we then fitted the positions of each spine with a quadratic function. For this, we used Bayesian linear regression of a quadratic polynomial function \citep[e.g.,][]{Bishop2006}. 

\begin{figure*}[tbp]
\centering
\includegraphics[width=\linewidth]{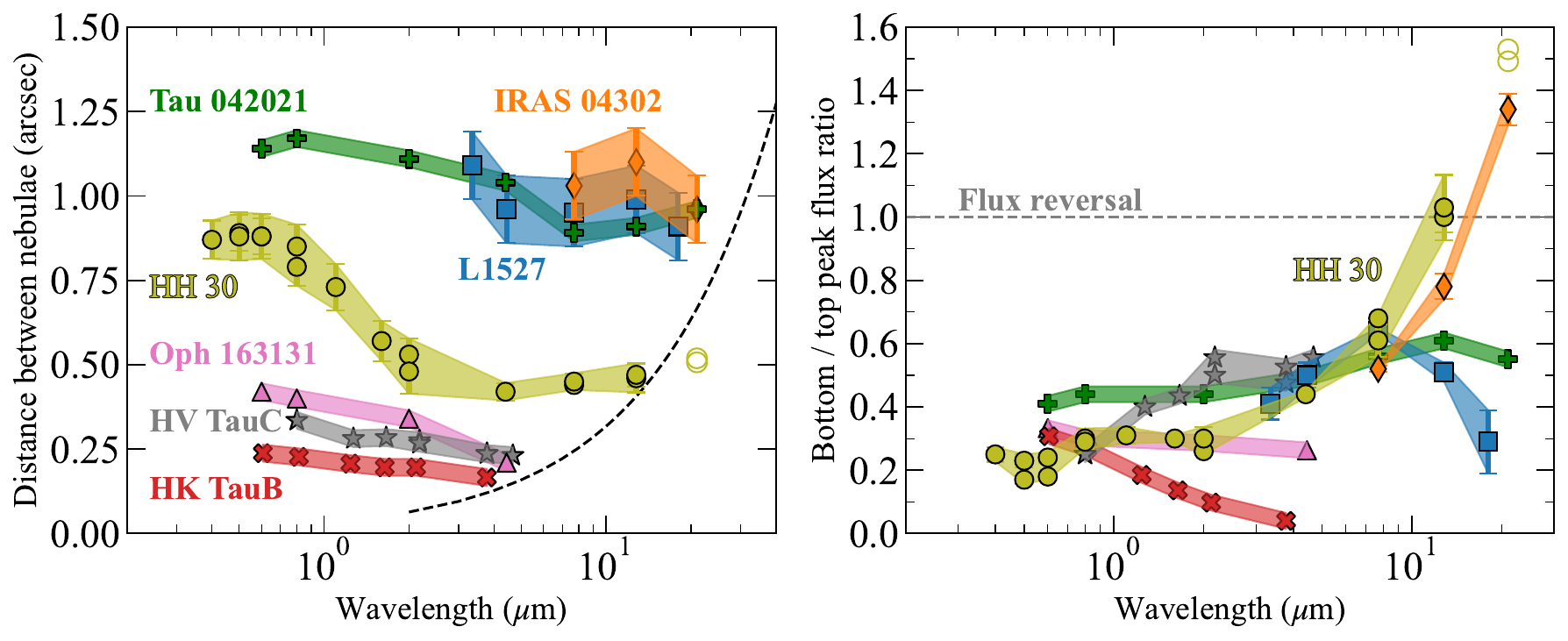}
\caption{Compilation of the nebular separation (left) and the bottom/top flux ratio (right) of various edge-on disks, including all our JWST/Cycle 1 targets (Tau042021, IRAS 04302, Oph163131, HH 30). Distance to all targets shown here lie in a similar range of 130-161 pc. In the left panel, the black dashed line is the FWHM of JWST's PSF ($\lambda/D$ with $D$ being the diameter of the telescope). The open symbols at 21 \micron~represent the estimation with PSF fitting. References: Tau 042021 \citep{Duchene24}, IRAS 04302 and L1527 \citep{Villenave24}, Oph 163131 \citep{Villenave_Oph163131}, HH 30 (this work), HV Tau C \citep{Duchene10}, HK Tau B \citep{McCabe11}. 
}
\label{fig:dwidth}
\end{figure*}

We did not apply the method described above for the 21-\micron~images because the disk is only barely resolved. However, we found that the central source has an FWHM perpendicular to the disk plane of $\sim0\farcs94$, whereas the corresponding FWHM of the WebbPSF \citep{Perrin14}\footnote{\url{https://jwst-docs.stsci.edu/jwst-mid-infrared-instrument/miri-performance/miri-point-spread-functions\#gsc.tab=0}} has $0\farcs674$\footnote{Note that WebbPSF assumes either a flat weighting or a G2V star as a source spectrum by default when calculating relative weighting across the bandpass. We tested the effect of the source spectrum by changing it from a default value to one that mimics the SED of HH 30 and found that the relative error of FWHM due to different source spectra is less than 3\%. It is, therefore, not critical for our discussion. }.
This shows that the emission is not a point source, but an extended emission.
We estimated the intrinsic size of the 21-\micron~emission by performing PSF fitting. We did a vertical cut passing through the center of the observed image and then fitted that intensity profile with two WebbPSF profiles. The position offsets and relative amplitudes of the two WebbPSF profiles were the parameters to fit the observed profile and were found with a least squares method. 
Table \ref{tab:mor} summarizes the extracted morphological properties of the disk reflection nebulae. The nebular separation is measured at the closest separation, and the flux ratio represents the peak-to-peak ratio of the two surfaces. Figure \ref{fig:dwidth} shows the nebular separations and the flux ratios of the two surfaces at various wavelengths, along with the results for other edge-on disks. 

The way the nebular separation shrinks with wavelength gives information on the vertical settling of dust grains as well as the opacity law of grains in the surface regions \citep{DAlessio01,DAlessio06, Watson04, Duchene24}. 
The nebular separation of the HH 30 disk decreases greatly from 0.4~\micron~to 2~\micron~\citep[see also][]{Cotera01, Watson04}. However, contrary to what was naively anticipated by extrapolating the HST results to mid-IR wavelengths, the separation remains nearly the same from 4~\micron~to 21~\micron. 
Since the nebular separation is larger than the PSF size of the telescope up to 12.8 \micron, the HH 30 disk is sufficiently geometrically thick  to be resolved with JWST at mid-IR wavelengths.

\begin{figure}[tbp]
\centering
\includegraphics[width=\linewidth]{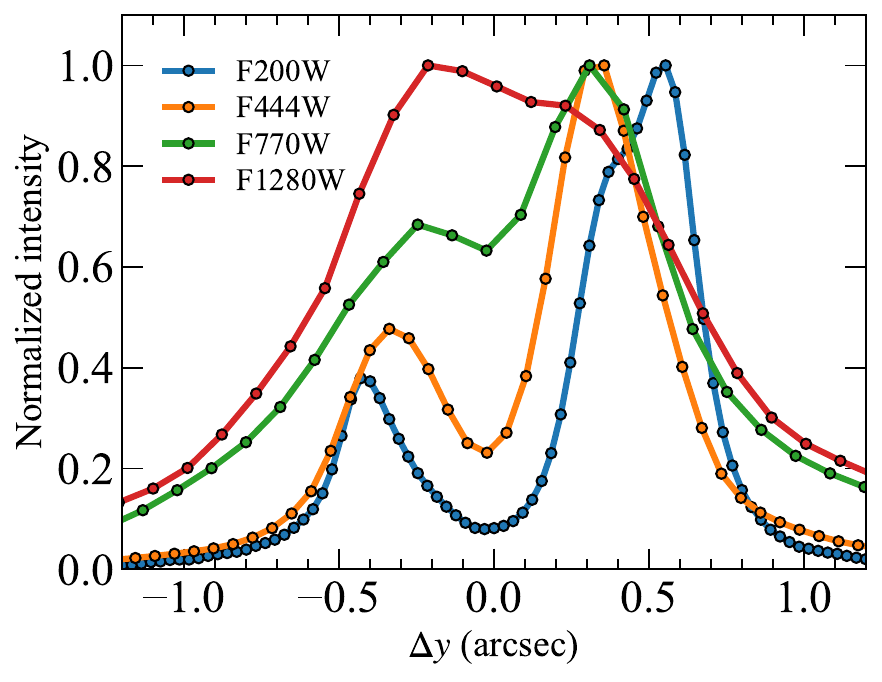}
\caption{Verical intensity profiles of the JWST September datasets measured at $\Delta x=0.8$\arcsec~from the symmetry axis (to northeast) to avoid contamination of the jet. The JWST curves are well aligned relative to each other by using a common field star seen in the images. }
\label{fig:vcut}
\end{figure}

Figure \ref{fig:dwidth} also reveals an interesting trend in the flux ratio of the two surfaces measured at which the nebular separation becomes closest. We found that the disk's bottom/top flux ratio varies significantly with wavelength, particularly at mid-IR.
The ratio is $\lesssim0.3$ for $\lambda\le2~\micron$, meaning that the top surface is brighter than the bottom surface by a factor of $\sim3.3$. 
However, the ratio increases for $\lambda>2~\micron$ and eventually reaches nearly unity at 12.8~\micron, that is, the two disk surfaces are equally bright in the mid-IR.
One might argue that at $\lambda=12.8~\micron$ the jet is very bright and may affect the flux ratio. To investigate the effect of the jet, Figure \ref{fig:vcut} shows the vertical intensity profiles at a location offset from the jet axis by $0\farcs8$ at which the contribution of the jet is considerably reduced. Even in this case, we found a nearly equal or even reversed flux ratio, suggesting that the jet contamination alone cannot account for the near-equal flux ratio at $12.8~\micron$. Furthermore, the flux ratio suggested by the PSF fitting of the 21-\micron~image also hints at a flux reversal.

An unequal brightness between the two surfaces is usually interpreted as evidence of the disk not being exact edge-on as a consequence of forward (and multiple) scattering. However, as we will show in Section \ref{sec:incl}, such an interpretation fails to explain the observed trend. So far, only two other edge-on disks are known to show the flux reversal: Flying saucer \citep{Grosso03} and IRAS 04302 \citep{Villenave24}. The trend seen in HH 30 closely follows the one seen in IRAS 04302. We will further discuss some possible origins for the wavelength dependence in Section \ref{sec:fluxinversion}. 

\begin{table*}[ht]
    \centering
    \begin{threeparttable}
    \caption{Disk Morphological Properties of HH 30}
    \label{tab:mor}
    \begin{tabularx}{\linewidth}{lllcccccc}
        \toprule
        Instrument & Filter & $\lambda$ ($\mu$m) & Date & $d_{\text{neb}}$ (\arcsec) & Flux Ratio & FWHM$_{\text{top}}$ (\arcsec) & FWHM$_{\text{bot}}$ (\arcsec) & FW10\%$_{\text{top}}$ (\arcsec) \\
        \hline
 HST/WFPC2$^a$ & F439W & 0.40 & 1998/12/01 & $0.87 \pm 0.057$ & $0.25 \pm 0.02$ & $0.66 \pm 0.11$ & $0.72 \pm 0.05$ & $1.87 \pm 0.17$ \\
 HST/WFPC2$^a$ & F555W & 0.50 & 1998/12/01 & $0.89 \pm 0.053$ & $0.23 \pm 0.02$ & $0.72 \pm 0.17$ & $0.71 \pm 0.05$ & $2.03 \pm 0.13$ \\
 HST/WFPC2$^a$ & F555W & 0.50 & 2008/11/10 & $0.88 \pm 0.071$ & $0.17 \pm 0.01$ & $0.28 \pm 0.05$ & $0.85 \pm 0.07$ & $2.22 \pm 0.05$ \\
 HST/WFPC2$^a$ & F675W & 0.60 & 1998/12/01 & $0.88 \pm 0.067$ & $0.18 \pm 0.01$ & $0.70 \pm 0.05$ & $0.77 \pm 0.05$ & $2.08 \pm 0.05$ \\
 HST/WFPC2$^a$ & F675W & 0.60 & 2008/11/10 & $0.88 \pm 0.053$ & $0.24 \pm 0.02$ & $0.22 \pm 0.05$ & $0.77 \pm 0.05$ & $1.94 \pm 0.05$ \\
 HST/WFPC2 & F814W & 0.80 & 1998/12/01 & $0.85 \pm 0.065$ & $0.30 \pm 0.03$ & $1.07 \pm 0.18$ & $0.78 \pm 0.05$ & $2.30 \pm 0.11$ \\
 HST/WFPC2 & F814W & 0.80 & 2008/11/10 & $0.79 \pm 0.056$ & $0.29 \pm 0.01$ & $1.45 \pm 0.39$ & $0.95 \pm 0.05$ & $2.72 \pm 0.09$ \\
 HST/NICMOS & F110W & 1.1 & 1997/09/29 & $0.73 \pm 0.068$ & $0.31 \pm 0.03$ & $1.17 \pm 0.05$ & $1.14 \pm 0.07$ & $2.48 \pm 0.09$ \\
 HST/NICMOS & F160W & 1.6 & 1997/09/29 & $0.57 \pm 0.059$ & $0.30 \pm 0.01$ & $1.10 \pm 0.05$ & $1.09 \pm 0.08$ & $2.44 \pm 0.07$ \\
 HST/NICMOS & F204M & 2.0 & 1997/09/29 & $0.53 \pm 0.048$ & $0.26 \pm 0.01$ & $0.95 \pm 0.05$ & $0.88 \pm 0.05$ & $2.36 \pm 0.05$ \\
 JWST/NIRCam & F200W & 2.0 & 2023/09/07 & $0.48 \pm 0.066$ & $0.30 \pm 0.04$ & $0.72 \pm 0.09$ & $0.67 \pm 0.10$ & $2.12 \pm 0.05$ \\
 JWST/NIRCam & F444W & 4.4 & 2023/09/07 & $0.42 \pm 0.026$ & $0.44 \pm 0.01$ & $0.59 \pm 0.05$ & $0.52 \pm 0.05$ & $1.53 \pm 0.05$ \\
 JWST/MIRI$^b$ & F770W & 7.7 & 2023/01/23 & $0.44 \pm 0.014$ & $0.68 \pm 0.01$ & $0.72 \pm 0.05$ & $0.65 \pm 0.05$ & $1.50 \pm 0.05$ \\
 JWST/MIRI$^b$ & F770W & 7.7 & 2023/09/07 & $0.45 \pm 0.025$ & $0.61 \pm 0.05$ & $0.69 \pm 0.05$ & $0.67 \pm 0.06$ & $1.47 \pm 0.05$ \\
 JWST/MIRI$^{a,b}$ & F1280W & 12.8 & 2023/01/23 & $0.46 \pm 0.042$ & $1.00 \pm 0.05$ & $0.70 \pm 0.06$ & $0.76 \pm 0.05$ & $1.56 \pm 0.05$ \\
 JWST/MIRI$^{a,b}$ & F1280W & 12.8 & 2023/09/07 & $0.47 \pm 0.034$ & $1.03 \pm 0.10$ & $0.70 \pm 0.05$ & $0.73 \pm 0.05$ & $1.67 \pm 0.05$ \\
 JWST/MIRI$^c$ & F2100W & 21 & 2023/01/23 & $\sim0.51$ & $\sim1.49$ & - & - & - \\
 JWST/MIRI$^c$ & F2100W & 21 & 2023/09/07 & $\sim0.52$ & $\sim1.53$ & - & - & - \\
\hline
    \end{tabularx}   
    \begin{tablenotes}[flushleft]
    \item \raggedright {\textbf{Notes.} The nebular separation, $d_{\text{neb}}$, is the closest distance between the parabola to each spine, and their error bars correspond to $1\sigma$ confidence intervals of the spine fitting. The bottom/top flux ratio is measured along the vertical cut at the closest separation.
    FWHM$_{\text{top}}$ and FWHM$_{\text{bot}}$ represent the full-width half maximum of the lateral intensity profile along the top and bottom spines, respectively, while FW10\%$_{\text{top}}$ is the 10\% width for the top spine. 
    The error bars of the flux ratio and FWHMs were estimated based on the region with $1\sigma$ confidence intervals of the spine fitting or 0.01 for the flux ratio and 0\farcs05 for FWHMs, whichever is greater. The low values for $\mathrm{FWHM}_\mathrm{top}$ of F555W and F675W taken in 2008 are due to a strong contamination of the jet.
    \item \raggedright $^a$
    Data set with significant contamination by line emission along the collimated jet.
    \item \raggedright $^b$ Nebula separation was measured by means of two-component Gaussian fitting.
    \item \raggedright $^c$ Estimated based on the PSF fitting of the vertical intensity profile.
    }
    \end{tablenotes}
    \end{threeparttable}
\end{table*}


\subsubsection{A conical outflow} \label{sec:conical}

The NIRCam/2-\micron~image reveals a conical structure as well as the collimated jet extending vertically up to $\sim3$\arcsec~from the disk midplane (Figure \ref{fig:nircam}). The observed structure is reminiscent of HST NICMOS images of a Class I protostar DG Tau B \citep{Padgett99}, where the conical structure is due to line emission of H$_2$ in the outflow \citep{Delabrosse24}. 
Recently, \citet{Pascucci24} found in their NIRSpec/IFU observations of HH 30 that H$_2$ emission shows a conical structure similar to the one seen in our NIRCam/2-\micron~image, whereas their dust continuum (scattered light) image does not reveal such a structure. Hence, the conical structure seen in our image is likely due to H$_2$ line emissions in the bandpass rather than scattered light by dust grains.  

\begin{figure}[tbp]
\centering
\includegraphics[width=\linewidth]{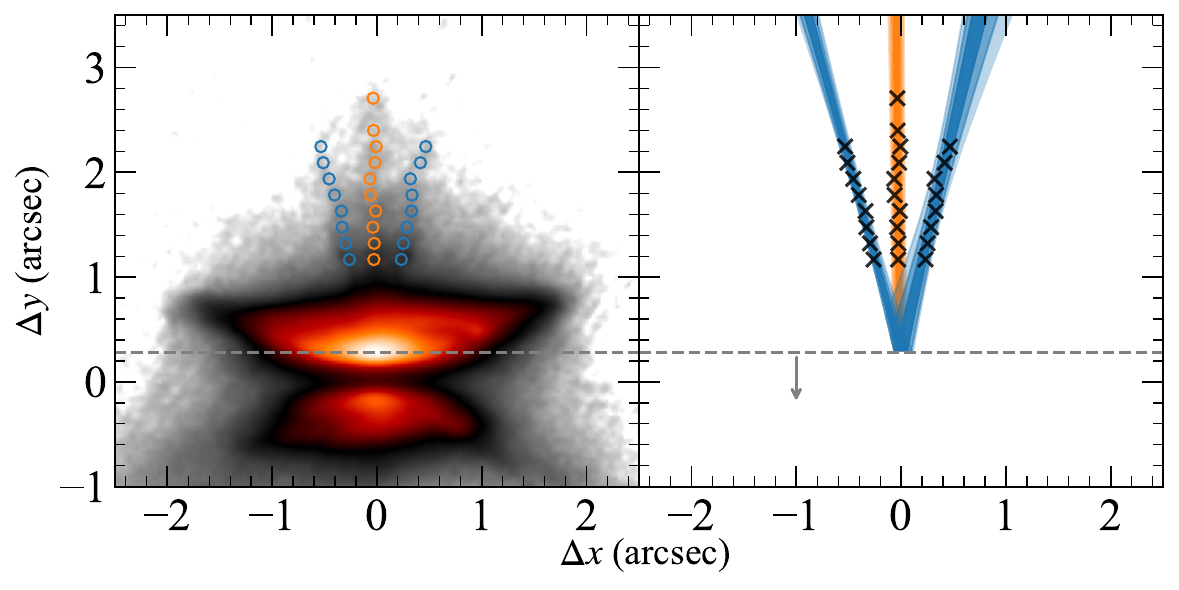}
\caption{Fitting of the conical structure and the jet. Left panel shows the extracted locations of the local intensity peaks. Right panel shows the results of the Bayesian linear regression for the jet and conical structure with a linear base function. The color gradation indicates the [1$\sigma$,2$\sigma$,3$\sigma$] confidence levels computed from the Bayesian regression. The central star is located below the horizontal dashed line. }
\label{fig:outflow}
\end{figure}

Here we estimate the opening angle and the launching radius of the conical outflow and the jet orientation.
First, we made a transverse cut of the image by changing the altitude from the disk midplane.
To improve the signal-to-noise ratio, we binned each cut with a five-pixel width ($\approx0.15\arcsec$) and identified the three intensity peaks (two at the sides and one at the center) using three-component Gaussian functions. 
The peak positions are then fitted with a linear function with Bayesian regression to derive the opening angle and the launching location, as shown in Figure \ref{fig:outflow}. 
As a result, we found that the jet PA (w.r.t. North) is $32.2^\circ\pm0.4^\circ$, and the left and right sides of the conical outflow have a semi-opening angle of $14.7^\circ\pm1.0^\circ$ and $13.5^\circ\pm1.8^\circ$ measured from the jet axis, respectively. 
The values are consistent with the semi-opening angle of H$_2$ emission found in NIRSpec/IFU observations by \citet{Pascucci24}: $14.1^\circ\pm0.5^\circ$ (left side) and $12.8^\circ\pm0.5^\circ$ (right side).
In contrast, the semi-opening angle of $^{12}$CO outflow of HH 30 measured with ALMA is $\sim35^\circ$ \citep{Louvet18} and it is wider than the values found in near-IR observations, suggesting a nested structure of the conical outflow \citep[see also][]{Pascucci24}. The inferred jet PA is slightly higher than 31.6$^\circ$ reported in \citet{Anglada07}, although our value is still consistent with a jet PA of $32^\circ\pm2^\circ$ measured in \citet{Pascucci24}.

We estimated the launching radius of the conical outflow in Figure \ref{fig:outflow} by simply assuming that the trajectory of the outflow can be expressed with a linear function down to the midplane.
Since we do not know the exact stellar location in the image plane, we assume that the central star is located below the brightest pixel in the scattered light image, as shown with the dashed line in Figure \ref{fig:outflow}. 
From the intersection of the fitted line and the dashed line, we obtained an upper limit on the launching radius of $\sim 16$ au by taking a $3\sigma$ confidence interval. 
The derived launching radius is in agreement with the launching radius of $^{12}$CO outflow ($\le22$ au) derived in \citet{Louvet18}.

\subsubsection{Spatially resolved jet at 12.8~\micron} \label{sec:jet}
\begin{figure}[tbp]
\centering
\includegraphics[width=0.34\linewidth,angle=-90]{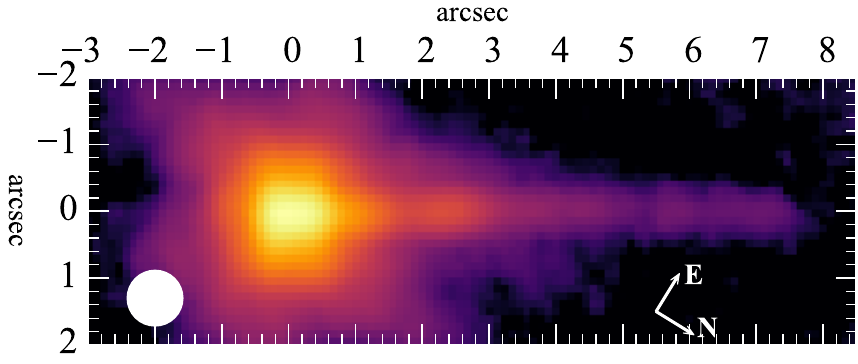}
\caption{
The MIRI/F1280W (12.8 \micron) image taken in January 2023 with the jet orientated horizontally with labels in units of arcsec. The size of the PSF is shown in the bottom left of the image.}
\label{fig:jet}
\end{figure}

\begin{figure*}[tbp]
\centering
\includegraphics[width=\linewidth]{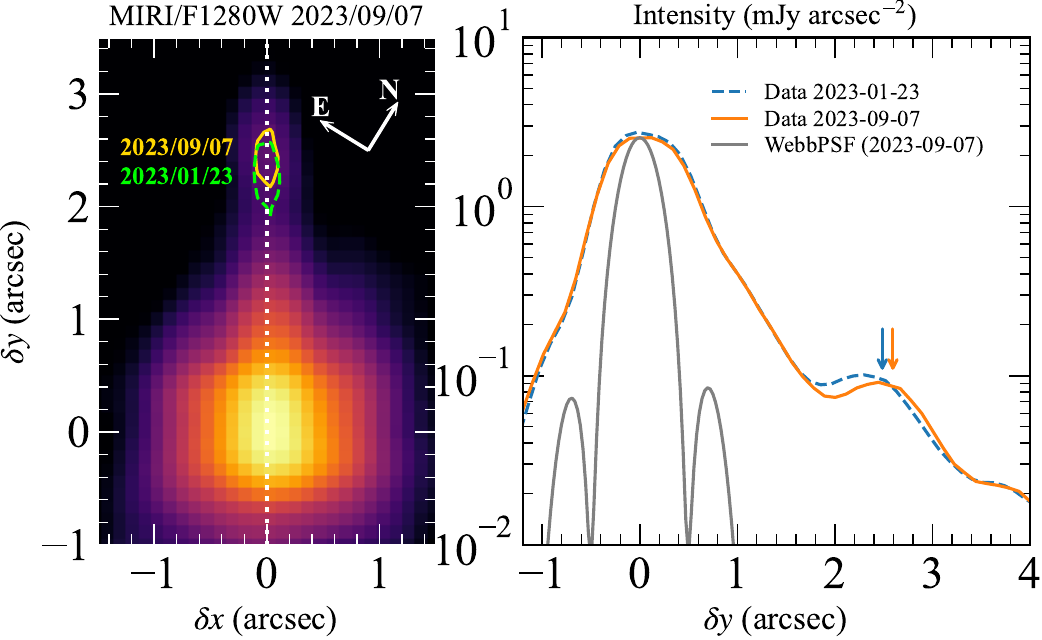}
\caption{(Left) Movement of a jet knot seen in MIRI/F1280W image between two epochs. The contours are the value at $10^{-1.5}$ relative to the peak. (Right) Vertical intensity profile of the F1280W images measured along the jet, as indicated with the white dotted line on the left panel. 
The gray line shows the corresponding PSF computed from WebbPSF.
}
\label{fig:knot}
\end{figure*}

Figure \ref{fig:jet} shows a wider field of view image of the mid-IR jet we detected in the MIRI/F1280W (12.8 \micron) images. The jet extends up to $\sim 8\arcsec$ from the disk midplane. Unlike the optical jet seen in the HST image (e.g., Figure \ref{fig:rgb}), the counter jet is not visible in the MIRI images.
Although our F1280W image does not contain spectroscopic information, a Spitzer/IRS spectrum of HH 30 retrieved from the Combined Atlas of Sources with Spitzer IRS spectra (CASSIS) \footnote{\url{https://cassis.sirtf.com/atlas/welcome.shtml}} \citep{Lebouteiller11} confirms the presence of a strong atomic line of [NeII] 12.8~\micron~in the bandpass, suggesting the observed mid-IR jet to be attributable to it. 

We also detected for the first time a proper motion of an emission knot associated with the mid-IR jet between January 2023 and September 2023, as shown in Figure \ref{fig:knot}. The two images are manually aligned by eye based on the disk emission. 
The intensity profile along the jet axis clearly shows the proper motion of a knot, which is located $\sim2\farcs5$ above the disk midplane.
To estimate the apparent velocity of the knot, we first fit the intensity profile along the dotted line shown in Figure \ref{fig:knot} for $1\arcsec\le\delta y\le4\arcsec$, but excluding the knot with a third-order polynomial function to estimate the continuum level. We then subtracted it and applied a Gaussian fit to the resulting continuum-subtracted curves to derive the central location of the knot.
As a result, we found the knot moves 0\farcs109 between the two epochs, which is then translated into a velocity of 0\farcs174/yr. 
By assuming a distance $d=146.4$ pc, the knot is moving with a 
velocity of $121$ km s$^{-1}$.
This velocity is in agreement with a typical proper motion velocity of the optical emission knots, $\sim100$ km s$^{-1}$ \citep{Estalella12}.

\subsection{Millimeter dust continuum} 

\label{sec:alma}
\begin{figure*}[t]
\begin{center}
\includegraphics[width=\linewidth]{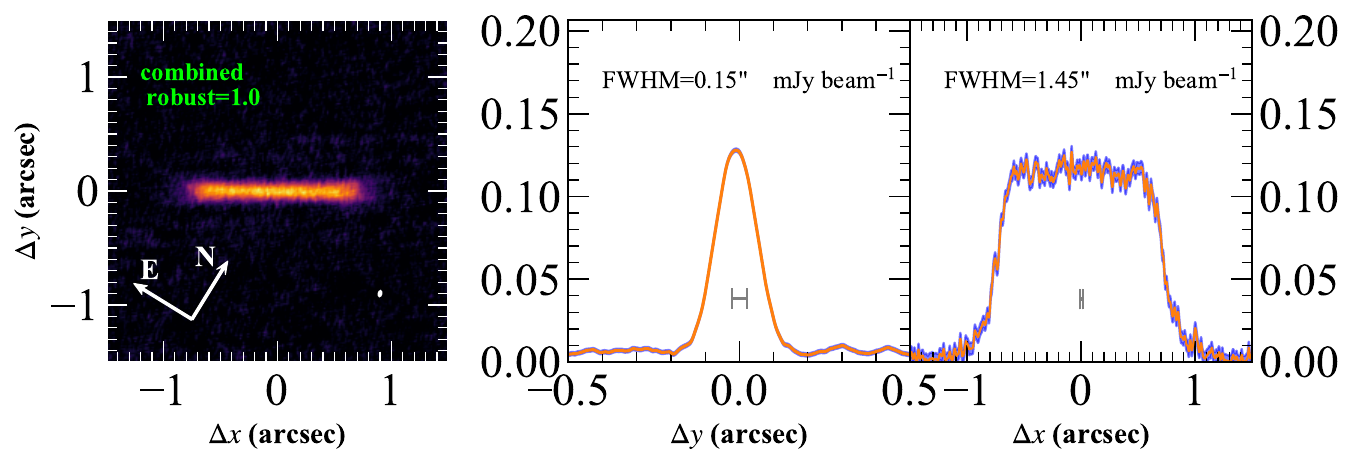}
\includegraphics[width=\linewidth]{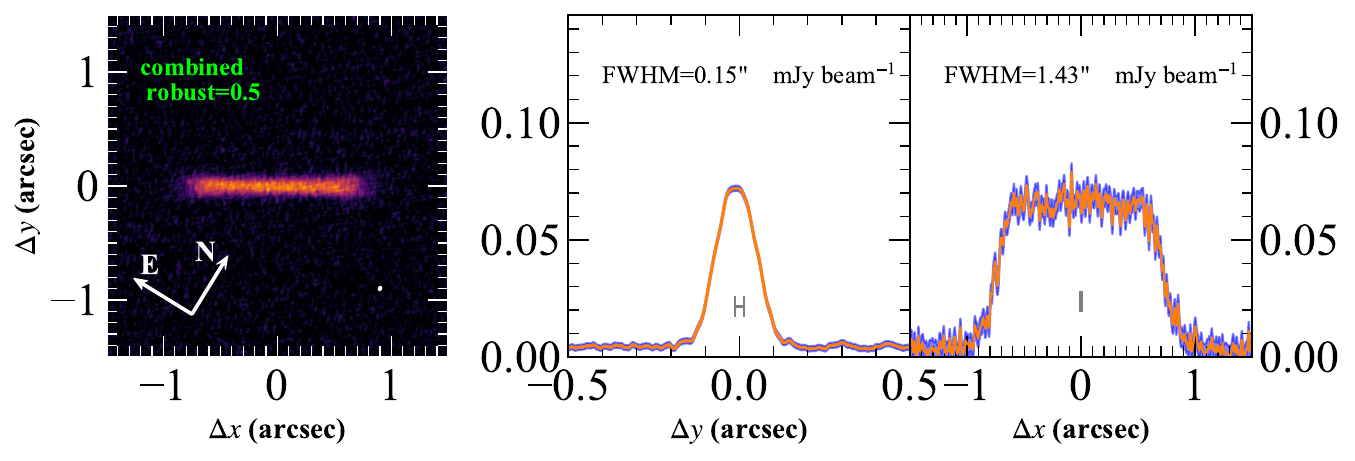}
\caption{Top row: (Left) ALMA Band 6 image of the dust continuum emission of HH 30 with a linear stretch from zero to the peak value. We combined the extended and compact configurations with a robust parameter of 1.0, and the corresponding beam size is $0\farcs046\times0\farcs023$, as shown in white at bottom right. (Middle) The vertical intensity profile averaged within $\Delta x=\pm0\farcs73$. (Right) The horizontal intensity profile averaged within $\Delta y=\pm0\farcs06$. In the middle and right panels, the gray marker indicates the beam size along each intensity cut. The blue-shaded region corresponds to the $3\sigma$ errors. Bottom row: Same as the top row, but for a robust parameter of 0.5 with a beam size of 0.029$\times$0.02\arcsec.}
\label{fig:almaimage}
\end{center}
\end{figure*}

\begin{figure}[t]
\begin{center}
\includegraphics[width=\linewidth]{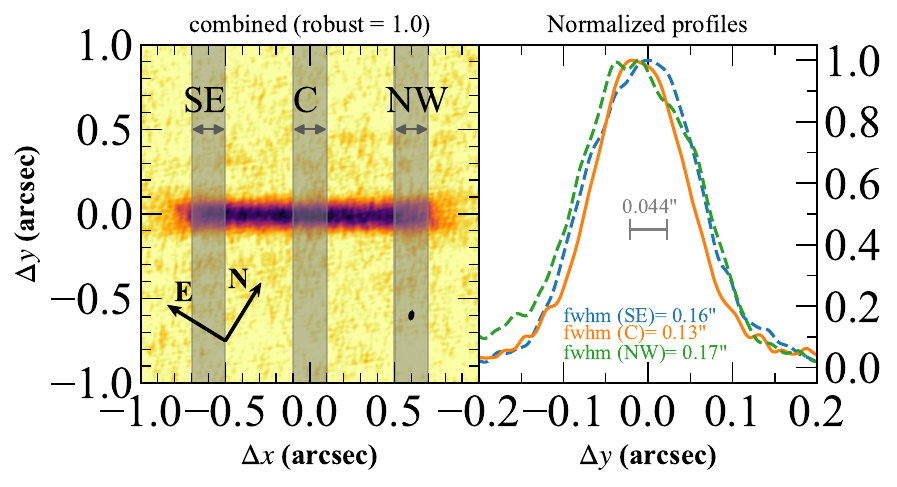}
\caption{The vertical thickness of the disk at three different locations (SE: southeast, C: center, and NW: northwest) for the ALMA images with a robust parameter of 1.0.  The gray marker in the middle of each panel shows the beam size. }
\label{fig:almaheight}
\end{center}
\end{figure}

Figure \ref{fig:almaimage} shows the dust continuum emission at the ALMA Band 6, where both compact and extended configurations are combined with a robust parameter of 1.0 (top) and 0.5 (bottom). 
The vertical and horizontal dust emission are clearly spatially resolved.
The image shows that the morphology of the emission is boxy.
Despite the fact that the vertical profile is spatially resolved, we do not find evidence of a dark lane or a shoulder peak.
Also, the horizontal profile shows a fairly uniform brightness inside $0\farcs6$, and there is no hint of a centrally peaked emission nor an inner dust cavity, supporting the findings by \citet{Louvet18}.
We measure the disk size using the 5$\sigma$ contour encompassing the disk emission for the robust=1.0 image. The full thickness and diameter of the disk are found to be 0\farcs22 and 1\farcs6, respectively. 
The aspect ratio of the dust continuum emission is about $0.14$, although this value does not represent an aspect ratio for the dust scale height, as the apparent disk thickness depends also on the optical depth of the disk (see Section \ref{sec:mmmodel} for an estimation of the dust scale height).
Also, we find that the disk appears to be narrower in the central region than in the side regions, as shown in Figure \ref{fig:almaheight}. 
The FWHM of the vertical profiles is 0\farcs13 at the central region, whereas it is 0\farcs16--0\farcs17 close to the disk edges. We also confirmed that the same trend appears in the image with a robust parameter 0.5, which has a better spatial resolution but with a reduced sensitivity. 

We measured the disk integrated flux within the  $3\sigma$ contour encompassing the HH 30 disk, where $\sigma$ is the rms of the continuum image. The integrated disk flux is $23\pm1.2$ mJy, 
where we assumed $5\%$ flux calibration error. 
The integrated flux is consistent with \citet{Louvet18}, where the authors measured 22.30 mJy at the ALMA Band 6. 

\section{Modeling} \label{sec:model}

We perform radiative transfer simulations in order to interpret the multiwavelength disk appearance revealed by HST, JWST, and ALMA. The goals of the modeling are to investigate (i)  the vertical settling of micron-sized grains and pebbles, (ii) the "extinction curve" of protoplanetary dust grains, and (iii) the inclination angle of the HH 30 disk. We do not attempt to perform exact image matching. Instead, we employ a simple disk model based on previous studies and see to what extent such a simple model can account for the observed morphology of the disk. 

\subsection{Star and disk models}

We assume a single star with an effective temperature of 3500 K in accordance with a spectral type of M0$\pm2$ \citep{White04}. 
Since the bolometric luminosity is unknown, we treat the bolometric luminosity (or the stellar radius) as a free parameter. The star is surrounded by an axisymmetric disk with a power-law dust surface density $\sigd\propto r^{-1}$, where $r$ is the distance from the star. The inner and outer radii are set as 2 au and 280 au, respectively. In the model calculations, we assume a distance to HH 30 to be 140 pc to make the comparison with previous modeling results easier \citep[e.g.,][]{Burrows96, Cotera01, Watson04, Madlener12}. 
The outer radius of 280 au will have $\approx 2\arcsec$, which is in agreement with the outer disk radius seen in the scattered light images (Figure \ref{fig:hstjwstalma}).  
Although little is known about the inner radius of the disk surrounding HH 30, our choice is in agreement with the inner truncation radius estimated by \citet{Louvet18}, where the authors found an upper limit of less than $7$ au based on the angular momentum of the CO molecular outflow. 
It seems possible that the inner radius is even closer in, down to a few stellar radii, as suggested from the presence of active accretion \citep{White04}. Although our choice was made for numerical convenience, this does not affect our conclusions.

The vertical distribution of dust grains is assumed to follow a Gaussian profile with the scale height of $\hd$, which is given by:
\begin{equation}
\hd(a)=\mathrm{max}\{\hfloor,\hg\mathrm{min}\{1,(a/\aset)^{-0.5}\}\},\label{eq:hd}
\end{equation}
where $\hg$ is the gas ($\approx$ small grains') scale height, $a$ is the grain radius, and $\aset$ is a free parameter to set the degree of dust settling and is defined as the maximum grain size that is fully vertically mixed.
In other words, small grains ($a\le\aset$) will have the same scale height as the gas, while large grains ($a\ge\aset$) will have a smaller vertical extent than the gas. 
For the gas scale height, we adopt $\hg=h_{100}(r/100~\au)^\beta$. 
We set $h_{100}=15~\au$ and $\beta=1.3$ in accordance with previous studies of scattered light modeling of HH 30 \citep{Burrows96, Cotera01, Watson04, Madlener12}. 
For numerical convenience, we set the floor value of the dust scale height as $\hfloor/r=0.01$. Therefore, if sufficiently large grains are present, a dust scale height can be as small as 1 au at $r=100$ au but not less than that. 
We note that the expression of Equation (\ref{eq:hd}) is not physically accurate in the sense that the dust scale height is solely given by the grain size rather than the Stokes number \citep{Dubrulle95, Youdin07}, which depends on a local gas density.
 
We also reduced the outer disk radius of large grains in our model to match the observed 1.3 mm dust continuum emission. We set the outer radius of grains with $a>10~\aset$ to be 100 au, motivated by the disk size in the ALMA image (Figure \ref{fig:almaimage}), while the outer radius of small grains $a<\aset$ is kept the same as the outer radius of the disk gas, i.e., 280 au. For $\aset\le a\le10~\aset$, the outer disk radius was determined by linear interpolation between the two cases. 

The total dust mass and the stellar luminosity are scaled so as to reproduce the dark lane thickness ($\approx0.5$\arcsec) and the integrated flux at the 2-$\mu$m disk image. In this study, we assume $i=89.9^\circ$ by default, as motivated from ALMA observations (Section \ref{sec:incl}).

\subsection{Radiative transfer simulations}

Given the star and disk model, we can calculate the dust temperature, spectral energy distribution (SED), and images by using a 3D Monte Carlo radiative transfer code \texttt{RADMC-3D} \citep{Dullemond12} with a full scattering matrix treatment. 
We used a 2D axisymmetric mode with the spherical coordinate system. In this model, we need to define radial and zenith-angle grids for the spatial cells. The radial grids are logarithmically equi-spaced with 256 cells from 2 to 300 au. The innermost radial cells are further radially divided into 10 sub-cells to avoid them being extremely optically thick. The polar grid cells are linearly spaced between $\pi/6$ and $5\pi/6$ with 512 grids. Each polar grid cell width is $\approx0.004$ radian, which is sufficient to resolve the vertical height of all grain sizes used in this study (i.e., the minimum aspect ratio of a dust layer is $\hfloor/r=0.01$).

We consider a power-law grain-size distribution obeying $n(a)da\propto a^{-3.5}da$ ($\amin\le a \le \amax$), where $n(a)da$ is the number density of grains within the grain radius range [$a$, $a+da$], and $\amin$ and $\amax$ are the minimum and maximum grain radii, respectively. 
In this study, we consider $\amin=10^{-2}$ \micron~to $\amax=10^4$~\micron.
We calculated their optical properties using 180 bins in a logarithmic space, which is sufficient to suppress the resonances.
We used the Distribution of Hollow Spheres method (DHS) \citep{Min05}, which can mimic the optical properties of compact dust aggregates \citep{Min16}, with an irregularity parameter $f_\mathrm{max}=0.8$. The grain composition and porosity are taken from the DIANA opacity model \citep{Woitke16}. The optical properties are calculated using a public code \texttt{OpTool} \citep{Dominik21}. 
Also, to implement the size-dependent vertical settling, we assumed that every 5 bins share the same dust scale height for a total of 180 grain size bins to reduce the computational cost.

The number of photon packages used for the thermal Monte Carlo and Scattering Monte Carlo simulations are $10^5$ and $3\times10^5$, respectively. 
Model images are convolved with a telescope PSF.
We used \texttt{TinyTim} \citep{Krist11} and \texttt{WebbPSF} \citep{Perrin14} to create PSFs of HST and JWST observations. For the ALMA image, we used a 2D Gaussian PSF with the beam size and orientation corresponding to our dataset.

\section{Modeling Results} \label{sec:modelres}

\subsection{Vertical dust settling in the HH 30 disk} \label{sec:sett}

\begin{figure*}[t]
\begin{center}
\includegraphics[width=\linewidth]{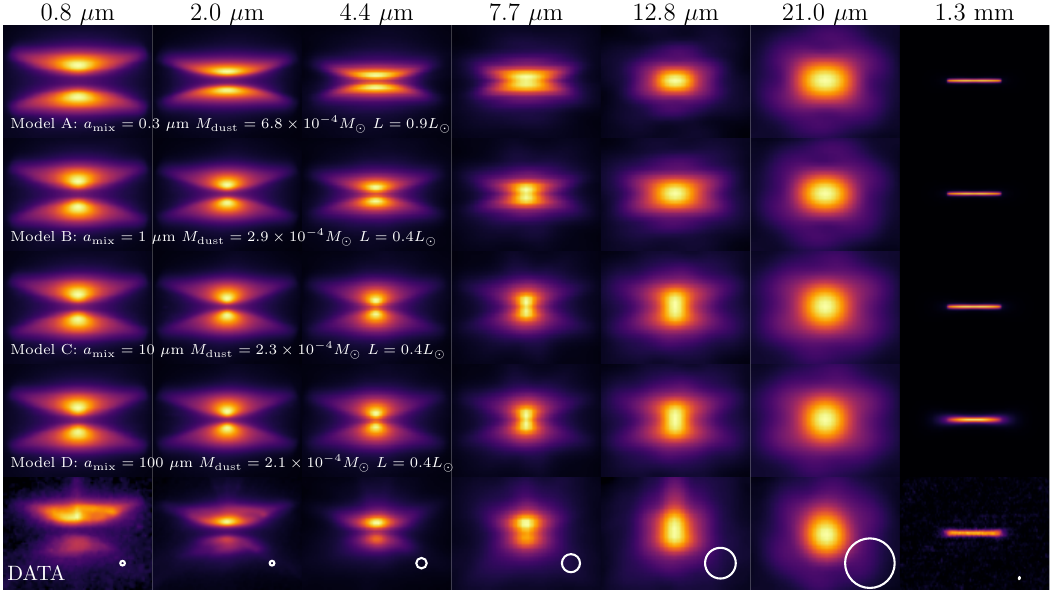}
\caption{Multiwavelength radiative transfer model images with a square root stretch from zero to the peak value. All images are shown with the same angular scale with the field of view of $4\arcsec\times4\arcsec$. The inclination angle is $89.9^\circ$. The models C and D appear nearly identical each other in the IR, but not at 1.3 mm. The total dust mass of each model was determined so as to reproduce the nebula separation at 2 \micron.}
\label{fig:modelimages}
\end{center}
\end{figure*}

Figure \ref{fig:modelimages} shows an overview of model images and observed images from $\lambda=0.8~\micron$ to 1.3 mm. 
The disk dust mass and stellar bolometric luminosity of each model were selected in a way that the 2-\micron~nebular separation and the integrated flux matches those of the observed image at this wavelength. This yields a disk dust mass of about $\approx$($2$--$7)\times10^{-4}M_\odot$ and a bolometric luminosity of $\approx0.4$--$1L_\sun$. 
In the following, we first discus the vertical settling of small grains based on the HST and JWST images and SED in Section \ref{sec:irmodel} and of large grains based on the ALMA image in Section \ref{sec:mmmodel}. 

\subsubsection{Settling of \micron-sized grains} \label{sec:irmodel}

\begin{figure*}[t]
\begin{center}
\includegraphics[width=\linewidth]{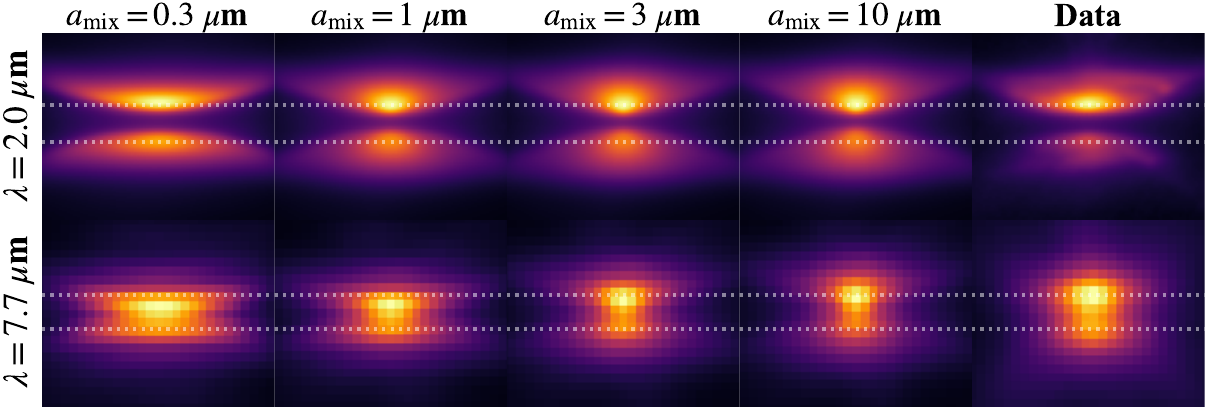}
\caption{Side-by-side comparison of 2-\micron~(top row) and 7.7-\micron~ (bottom row) images between models and the observations. Here we adopt an inclination angle of 88$^\circ$ to mimic the brightness contrast between the two surfaces, as seen in the observed MIRI image. 
The horizontal dashed lines are the locations of the top and bottom spine apexes in the observed image at each wavelength. Both model and observational images are shifted such that the apex of the bottom spine appears at the same height in the image.}
\label{fig:settimg}
\end{center}
\end{figure*}

\begin{figure*}[t]
\begin{center}
\includegraphics[width=\linewidth]{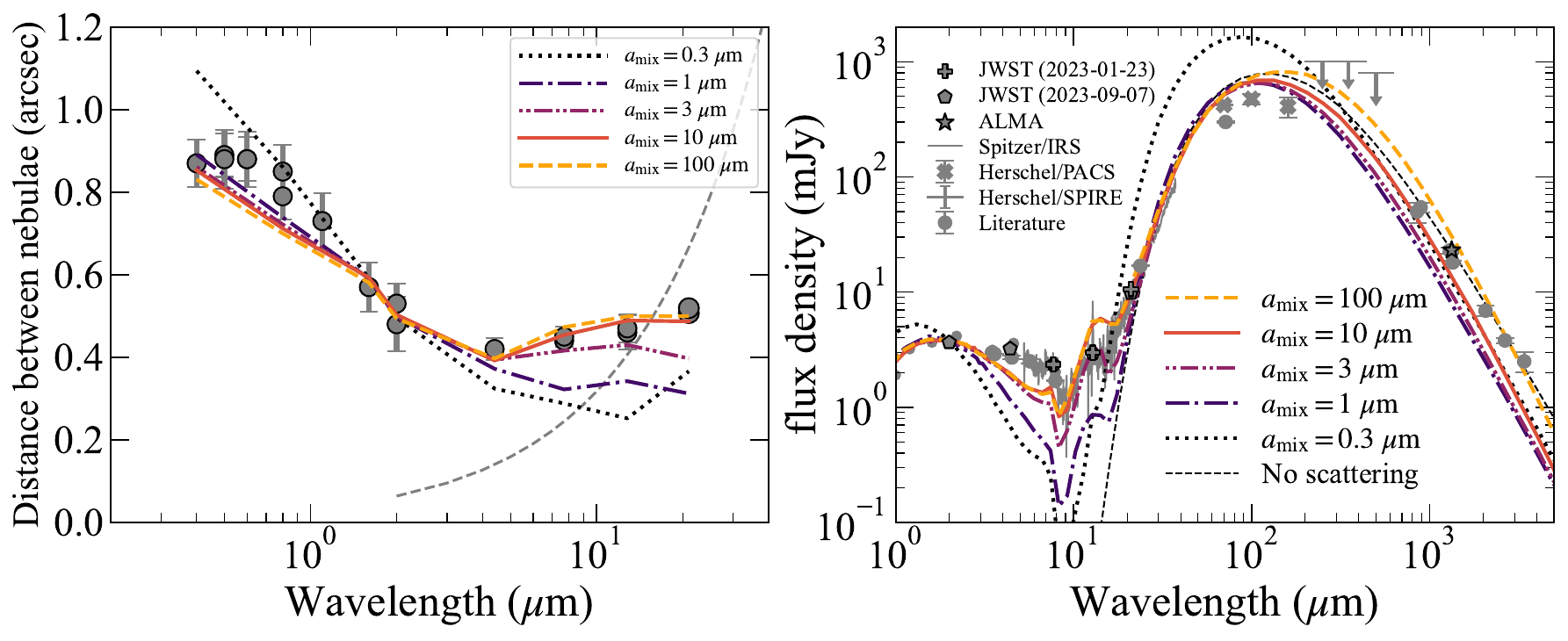}
\caption{(Left) The nebula separation as a function of wavelengths with various degrees of vertical dust settling. The inclination angle of the models is $89.9^\circ$, as in Figure \ref{fig:modelimages}. (Right) SED of HH 30. Different lines represent the SED of selected models with $\amax=1$ cm with different degrees of vertical dust settling. The dashed line shows the results without scattering for $\aset=10~\micron$. The photometric measurements below 2 \micron~were de-reddened to compensate for the interstellar extinction by assuming the estimated extinction of a region in which HH 30 is embedded $A_\mathrm{V}=3.3$ mag \citep{Dobashi05} using the extinction law of \citet{Cardelli89} with $R_\mathrm{V}=3.1$ mag. }
\label{fig:sett}
\end{center}
\end{figure*}

Figure \ref{fig:modelimages} shows that nebular separation at the optical and infrared wavelengths depends on $\aset$. As a general trend, a smaller value of $\aset$ leads to a narrower nebular separation at longer wavelengths, such as at 7.7 \micron.
To further clarify the impact of $\aset$ on the nebular separation,
Figure \ref{fig:settimg} shows 2-\micron~ and 7.7-\micron~ model images side by side. We also changed the inclination angle in Figure \ref{fig:settimg} to $88^\circ$ to mimic the observed flux ratio at 7.7~\micron.
By construction, all 2-\micron~ model images have a nebular separation consistent with the observation at this wavelength. However, at 7.7~\micron, the nebular separations for models with $\aset=0.3~\micron$ and $1~\micron$ are narrower than observed because larger grains are well settled in these models. As $\aset$ increases, larger grains are lifted up to the surface regions, and consequently, the nebular separation is maintained to longer wavelengths.
We found that the observed nebular separation is reproduced when $\aset\gtrsim3~\micron$. In Figure \ref{fig:sett}(left), we show the measured nebular separation as a function of wavelength. The observed wavelength dependence of the nebular separation is reasonably explained when $\aset\gtrsim3~\micron$. Increasing $\aset$ from 10~\micron~to an even larger size (i.e., 100~\micron~in our models) does not affect the appearance of the IR reflection nebulosity anymore. 
We also found that models with $\aset\gtrsim3~\micron$ show better an agreement with the observed SED of HH 30, as shown in Figure \ref{fig:sett}(right).
For the cases of $\aset=0.3~\micron$~and $1~\micron$, the model SEDs show a deep trough in the mid-IR, whereas for $\aset=3$~\micron~or larger, the mid-IR flux is significantly increased. Therefore, the observed appearance of the reflection nebulosity and its integrated flux in the mid-IR points to the presence of grains of about 3 \micron~in radius or larger in the disk surface.
We also confirmed that our dust model has an extinction opacity law that is consistent with the findings of previous studies (Appendix \ref{sec:opac}).

Figure \ref{fig:sett} also shows that the observed mid-IR emission is due to scattering, which is also the reason why the presence of micron-sized or larger grains in the disk surface is mandatory to produce high mid-IR fluxes.
In further support, switching off scattering in the models shows the SED to drop dramatically below $\sim20~\micron$, as shown in Figure \ref{fig:sett}. Therefore, the emissions in the F770W and F1280W filters must be dominated by scattered light.
The scattering efficiency of grains depends on how large the grains are relative to the wavelength, and it usually becomes significant when $2\pi a/\lambda\gtrsim1$ \citep{Bohren83}. 
At F1280W or $\lambda=12.8~\micron$, grains should be larger than $12.8/2\pi\simeq2$~\micron~to cause efficient light scattering in the mid-IR (see also Figure \ref{fig:opac}). 
To summarize, since micron-sized or larger grains cause an efficient mid-IR scattering, their presence at a higher disk latitude makes the disk appear bright and simultaneously geometrically thick, as observed in the MIRI images. 
  
Our simple model does not reproduce some of the observed properties. First, the models, especially for $\aset\ge3~\micron$, show strong forward scattering, which makes the lateral intensity distribution much more peaked than observed (see Figure \ref{fig:modelimages}). Models with smaller grains ($\aset=0.3$~\micron~or $1$~\micron) better reproduce the observed lateral intensity profile. Second, the models do not fully capture the nebular separation behavior between 0.8~\micron~and 2~\micron. Third, our model predicts too strong silicate absorption features in the SEDs. All of these issues would be solved or at least alleviated if we performed a finer parameter space study, exploring the parameters that we have fixed in our models, such as grain composition, grain shape/porosity, and grain size distribution. This is left for a future study.

\subsubsection{Settling of large grains} \label{sec:mmmodel}

\begin{figure}[t]
\begin{center}
\includegraphics[width=\linewidth]{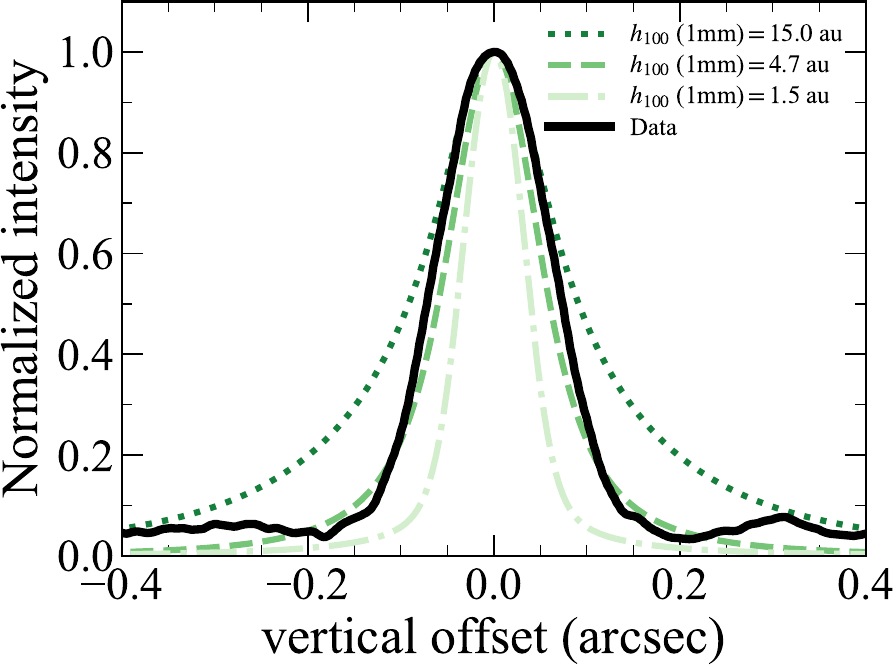}
\caption{The vertical intensity cut of the ALMA Band 6 image (robust=1.0). Dashed, solid, and dot-dashed lines represent the model with a dust scale height of mm-sized grains of 15 au (no settling), 4.7 au (model D), and 1.5 au (model C), respectively.}
\label{fig:almavcut}
\end{center}
\end{figure}
\begin{figure}[t]
\begin{center}
\includegraphics[width=\linewidth]{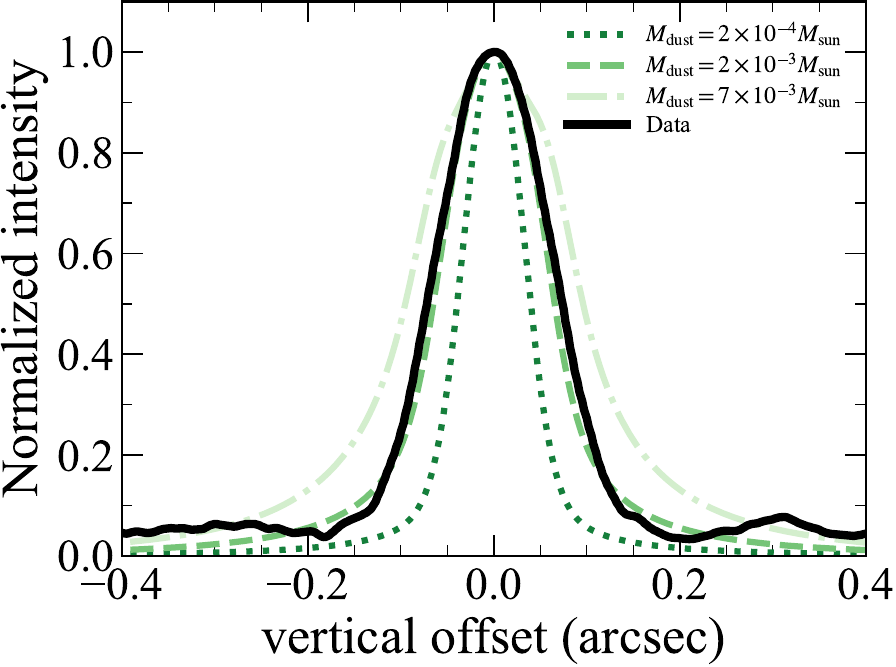}
\caption{Same as Figure \ref{fig:almavcut}, but for the increased disk mass for the model C, which has $h(a=1~\mathrm{mm})=1.5$ au at 100 au. }
\label{fig:almavcutmass}
\end{center}
\end{figure}

Next, we study the vertical settling of large grains based on the 1.3-mm image, which can affect the observed disk thickness.
To investigate this, 
Figure \ref{fig:almavcut} shows the brightness profiles along the minor axis of the disk for three models: $\aset=10~\micron$ (model C), $100~\micron$ (model D), and 1 cm (no settling) (see Figure \ref{fig:modelimages} for model names). These models have scale heights of mm-sized grains of 1.5 au, 4.7 au, and 15 au at 100 au from the central star, respectively. As the scale height becomes smaller, the profiles become thinner, and the model with $h_{100}=4.7$ au reproduces the observed brightness profile reasonably well, whereas the model with $h_{100}=1.5$ au does not. In contrast, the no-settling model produces a vertical profile thicker than the observations, illustrating the necessity of vertical settling of mm grains in the HH 30 disk.

The vertical brightness profile however also depends on the optical depth or the disk mass, and increasing disk mass could potentially explain the apparently broad brightness profile even if the scale height is small. To test this possibility, Figure \ref{fig:almavcutmass} shows the brightness profiles for model C, but with increased disk mass. It turns out that the observed vertical profile can only be explained if the disk {\it dust} mass is as high as $2\times10^{-3}M_\odot$. 
Assuming a gas-to-dust ratio of 100, the total disk mass would then be comparable to the stellar mass, which would be unusual for a Class II disk like HH 30. Following that argument, the dust scale height of mm grains in the HH 30 disk must be larger than 1.5 au at 100 au.

The inferred dust scale height of mm-sized grains is larger than that found in some highly settled Class II protoplanetary disks. In the HL Tau disk, \citet{Pinte16} found that the dust layer needs to be a geometrically thin layer: of the order of 1 au at 100 au. \citet{Villenave22} also found a highly settled dust disk around Oph 163131, whose scale height of mm-sized grains was constrained to 0.5 au at 100 au or less. \citet{Duchene24} also found in the disk around Tau 042021 that the scale height of the mm-grains is <1 au at 100 au. However, it is also suggested that younger disks have minimal or modest settling of mm-grains \citep{Villenave23, Villenave24, DanielLin23, Guerra24}. \citet{DanielLin23} found a scale height of mm grains in Class I IRAS 04302 disk to be $\sim6$ au at 100 au, which is comparable to the gas pressure scale height at that distance. Thus, the scale height of mm grains inferred for the HH 30 disk seems to be similar to such a younger disk, possibly implying a younger (or less evolved) nature for HH 30. 
The presence of a north-east large-scale nebula near HH 30 \citep{Burrows96} might also support this idea, although a further modeling is necessary to draw a robust conclusion.

\subsection{Inclination angle of the HH 30 disk} \label{sec:incl}

\begin{figure}[t]
\begin{center}
\includegraphics[width=\linewidth]{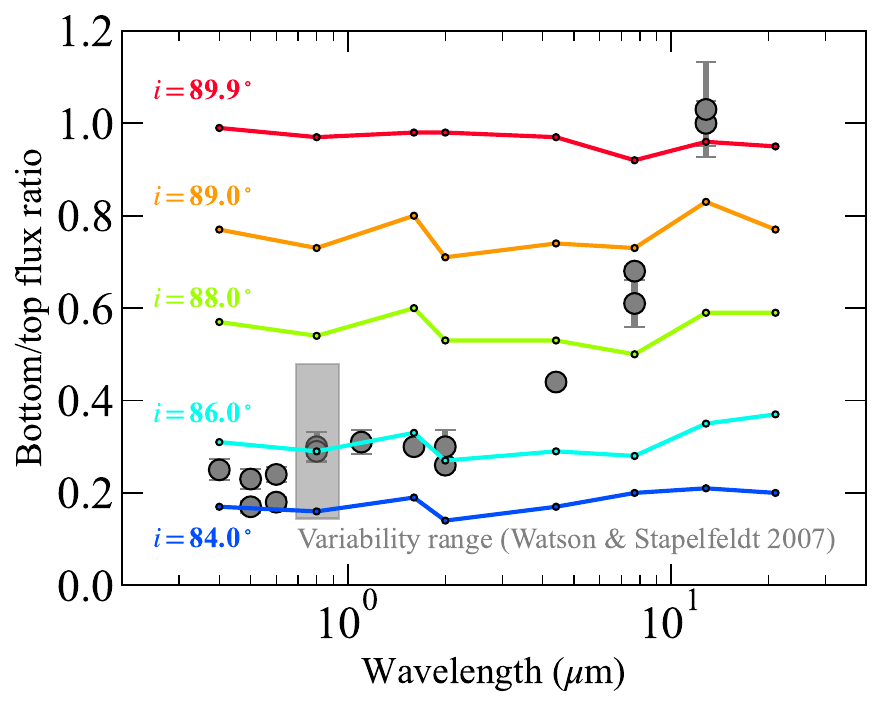}
\caption{Bottom/top surface flux ratio of the model C as a function of wavelength at various inclination angles. The gray box shows the range of time-varying flux ratio at HST/F814W \citep{Watson07}, although it is for an integrated flux ratio while our measurements are for a peak-to-peak ratio. The fluctuation of the values along each model curve is partly due to a different quality of fitting to the top and bottom spines, leading to the flux ratio being evaluated in slightly different vertical cuts from one wavelength to another. }
\label{fig:fluxratio}
\end{center}
\end{figure}

In Section \ref{sec:extmor}, we found that the flux ratio of the top and bottom nebulae significantly changes with wavelength. A common interpretation of the unequal fluxes between the two nebulae is due to the disk inclination angle being off the exact edge-on ($i=90^\circ$); however, it is unclear if this interpretation is capable of explaining the observed wavelength dependence.  
Here, with the goal to estimate HH 30's inclination angle, we investigate how the flux ratio of the model images varies with wavelength by changing the disk inclination angle.

Figure \ref{fig:fluxratio} shows the flux ratio from model C at various disk inclination angles. 
It turns out that the flux ratio in the optical/near-IR is well reproduced with $i=84^\circ$--$86^\circ$. 
However, the flux ratio in the mid-IR is only explained with a higher inclination angle, e.g., nearly $90^\circ$ at $\lambda=12.8~\micron$. 
One might wonder if the mismatch of the flux ratio between optical/near-IR and mid-IR may be due to the observed time variability of the HH 30 disk \citep{Burrows96, Watson07}. 
\citet{Watson07} analyzed 18 epochs of scattered light images of HH 30 from 1994 to 2005 and found that the (integrated) flux ratio between two surfaces varies from about 0.8 to 2.1 mag in F814W, corresponding to the range of the flux ratio from 0.48 to 0.14.
However, the NIRCam and MIRI images were taken in sequence with no time gaps, whereas the HST images in \citet{Watson07} were taken at scattered epochs spread over the years. The HST variability is all on much longer timescales than the 3 hour sequence of JWST observations taken in September 2023.  Comparing the nebula brightness ratio for F770W and F1280W between the two MIRI epochs shows no variability more than 2$\sigma$ (Table \ref{tab:mor}). Therefore, the change in flux ratio with wavelength from 2-21~$\mu$m is most likely intrinsic and not due to time variability.

\begin{figure*}[t]
\begin{center}
\includegraphics[width=\linewidth]{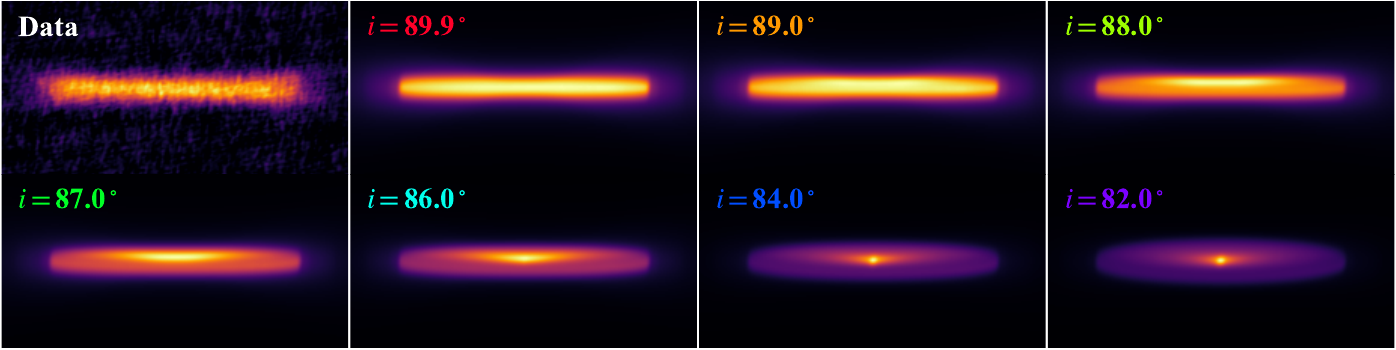}
\caption{Band 6 model images compared with the observation with a linear stretch from zero to the peak value. The inclination angle of the model disk is varied from $89.9^\circ$ to $82^\circ$. As the disk is inclined only by two degrees from the exact edge-on view, the brightness profiles along minor and/or major axes deviate from a uniform value.}
\label{fig:mmincl}
\end{center}
\end{figure*}

\begin{figure}[t]
\begin{center}
\includegraphics[width=\linewidth]{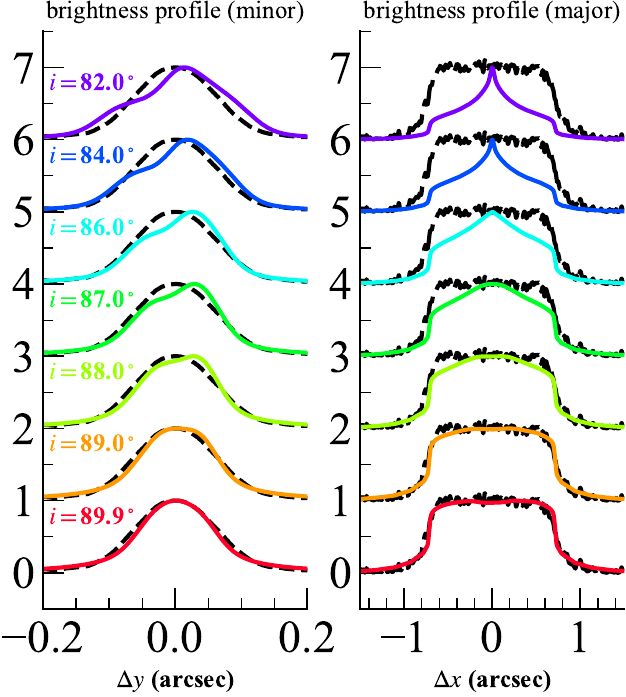}
\caption{Normalized brightness profiles of the images shown in Figure \ref{fig:mmincl} along the minor (left) and major axes (right) with vertical offsets for better visibility. The solid and dashed lines represent the model and observational results, respectively.}
\label{fig:mmprof}
\end{center}
\end{figure}

As we found a tendency for the disk to appear closer to exactly edge-on a longer wavelength, it is interesting to study what disk inclination would be required to explain the mm image.
We take the increased mass counterpart of model C, as introduced in Section \ref{sec:mmmodel} to study the geometry of the system. 
This model is physically unfavorable due to its high mass, but still useful to argue the geometry of the system as the model captures the normalized brightness profile both along the major and minor axes fairly well. 
Figure \ref{fig:mmincl} shows the model images at 1.3 mm at various inclination angles. We found that the disk appears rather boxy when it is close to the exact edge-on. However, as the inclination decreases, the central emission peak (due to the higher dust temperatures) starts to be revealed \citep[e.g.,][]{Villenave20}, which is not observed. Figure \ref{fig:mmprof} shows the normalized brightness profiles along the minor and major axes. It is clear that the observed profiles are only explained if $i\gtrsim89^\circ$. 
Therefore, the ALMA image also requires the disk to be very close to the edge-on.

Therefore, the multiwavelength images of the HH 30 disk cannot be characterized by a single value of the disk inclination angle, and a higher inclination angle is favored to reproduce the longer wavelength images.
This implies that the HH 30 disk potentially has an intricate disk geometry not captured by a smooth-flared disk model adopted in this study. We will discuss the potential origin in Section \ref{sec:fluxinversion}. 

\section{Discussion} \label{sec:discussion}

\subsection{Vertical settling in the HH 30 disk} \label{sec:uplift}

We found that grains of about 3 \micron~or larger appear to be fully mixed in the outer region of the HH 30 disk. This finding is consistent with previous SED modeling of HH 30 by \citet{Wood02}. The authors studied the SED of HH 30 using radiative transfer models and found that interstellar grains are so small that the SED shows a trough in the mid-IR range, similar to what we found for $\aset=0.3~\micron$ in Figure \ref{fig:sett}. They found that the models with larger grains, i.e., tens-of-microns, in the disk surface region can explain the observed SED. This conclusion is now supported not only by the SED, but also by the spatially resolved images of the disk at mid-IR wavelengths (Figure \ref{fig:settimg}). The 7.7-\micron~mid-IR images show that the upper surface extends radially up to $\sim50$--$100$ au, based on the FWHM$_\mathrm{top}$ and FW10\% in Table \ref{tab:mor} (with a distance of 146.4 pc). Thus, such large grains seem to be well mixed beyond 50 au from the central star. 

This is not the first time that the presence of >\micron-sized grains has been inferred on the surface of edge-on disks.
\citet{Pontoppidan07} found that a significant amount of 5--10~\micron-sized grains should be present in the outer surface region of the Flying Saucer edge-on disk beyond 50 au based on the SED analysis. 
\citet{Sturm23a} performed the SED analysis of the edge-on disk around HH 48 NE and found that the maximum grain size in the disk surface is about $\sim7~\mu$m at mid-IR wavelengths.
\citet{Sturm24} further tweaked their model to match the MIRI/MRS spectrum for the same target and the maximum grain size in the surface region is revised upward to $12.6~\micron$.
\citet{Duchene24} found evidence for fully mixed grains at least up to 10~\micron~in the edge-on disk around Tau 042021, as suggested by both the SED analysis and the spatially resolved disk images with JWST/MIRI (i.e., nebular separations in the mid-IR).  \citet{Villenave24} also found for IRAS 04302 that the disk images are consistent with well-mixed 10-\micron~grains. There are also some cases, where the presence of micron-sized grains in the disk surface is suggested for non-edge-on disks. For example, \citet{Mulders13} showed that the reddish and faint scattered light around HD 100546 is due to micron-sized grains in the disk surface. \citet{Tazaki21c} also found a similar reddish disk color in the disk around HD 142527 and attributed it to the 3-\micron~grains in the disk surface. 
Although there is growing evidence for the presence of larger grains in the disk surface, it is worth mentioning that there also exists a disk without a signature of large grains in the disk surface. The Oph 163131 disk is known as one of the most highly settled protoplanetary disk as revealed with ALMA \citep{Villenave22}. \citet{Villenave_Oph163131} studied this target with JWST/NIRCam and MIRI imaging observations and found that micron-sized grains are already well settled. Those observations clearly point to the diversity of the vertical settling in protoplanetary disks.

Micron-sized grains in the surface region are less coupled to the gas than sub-micron-sized grains, and therefore, they are more prone to settle down to the region where we cannot probe via IR observations. However, the vertical settling could be hindered by turbulent mixing \citep{Dubrulle95, Dullemond04}. 
We can estimate the vertical height at which grains decouple from the local gas and start to settle, as opposed to turbulent mixing. 
The settling starts when the Stokes number ($\mathrm{St}$) is comparable to the alpha parameter for the turbulent diffusion ($\alpha$). 
Denote $\zdec$ by the disk height at which $\mathrm{St}=\alpha$, it can be expressed as \citep{Binkert23}
\begin{equation}
\frac{\zdec}{\hg}=\sqrt{2\ln \frac{2\alpha\Sigma_\mathrm{g}}{\pi\rho_\bullet a}},\label{eq:zdec}
\end{equation}
where $\Sigma_\mathrm{g}$ is the gas surface density, $\rho_\bullet$ is the internal grain density, and $a$ is the grain radius. Equation (\ref{eq:zdec}) can be inverted to find $\alpha$ for a given value of $\zdec/\hg$:
\begin{equation}
\alpha = \frac{\pi \rho_\bullet a}{2\Sigma_\mathrm{g}}\exp\left[{\frac{1}{2}\left(\frac{\zdec}{\hg}\right)^2}\right] \label{eq:alpha}
\end{equation}
In order to explain mid-IR scattering in the HH 30 disk, grains need to exist above the $\tau=1$ surface at mid-IR wavelengths, i.e., 7.7~\micron. This requires $\zdec>z_\mathrm{\tau}$, where $z_\mathrm{\tau}$ is the height at which the optical depth measured from the central radiation source reaches unity. Model C has a gas surface density of $\sim1$ g cm$^{-2}$ at $r=100$ au from the central star (assuming a gas-to-dust ratio of 100), and we obtained $z_\mathrm{\tau}/\hg\sim 2$ in the mid-IR. By substituting those values together with the assumption of a compact spherical grain with a radius of  $a=3~\micron$ and a material density of $\rho_\bullet\sim2.8$ g cm$^{-3}$ (a material density of the DIANA grain composition, excluding the contribution of porosity), we obtain $\alpha\gtrsim10^{-2}$. 
If we consider porous grains, the internal density can be greatly reduced. 
Although the actual porosity of grains in the HH 30 disk remains greatly unknown, as an extreme case, we take a value of 99\% porosity, which corresponds to the case of a fractal aggregate made of sub-micron-sized monomer found in the disk around IM Lup \citep{Tazaki23}. In this case, $\rho_\bullet$ is reduced by a factor of $\sim100$, leading to $\alpha\gtrsim10^{-4}$. 
Despite having such an extreme value of dust porosity, the inferred $\alpha$ value is still higher than the values suggested in the disks around HL Tau and Oph 163131 \citep{Pinte16, Villenave20}, where $\alpha$ is inferred to be less than a few times $10^{-4}$.
Therefore, if the vertical dust distribution is set by settling/turbulent mixing balance, the $\alpha$ value of the HH 30 disk needs to be higher than those highly settled protoplanetary disks. Instead, the inferred moderate amount of settling in the HH 30 disk is reminiscent of less settled disks found in the younger evolutionary stage in Class 0/I sources \citep{Villenave23, Villenave24, DanielLin23, Guerra24} and a Class II disk \citep{Doi21, Doi23}.

\subsection{Inclination of the HH 30 disk} \label{sec:fluxinversion}

With a simple model, we found a tension in the disk inclination angle of the HH 30 disk inferred from our IR and mm images. The analysis suggests an inclination angle of $84^\circ$--$86^\circ$ for the optical/near-IR images, whereas $i\gtrsim89^\circ$ for the ALMA image. 
Our inclination inferred from the IR is consistent with the previous findings: $i=82^\circ\pm2^\circ$ \citep{Burrows96}, $i=84^\circ$ \citep{Cotera01, Wood02}, and $i=80.1^\circ$--$85.1^\circ$ \citep{Madlener12}. 
The value derived from our analysis is slightly higher than previously found, and this is likely because the scattering phase functions of our grain model are too forward-throwing compared to the observation, which likely lead to an overestimation of the inclination angle \citep[see also][]{Watson04}. 
The inclination angle inferred from the ALMA image is similar to the value inferred for the jet inclination of HH 30. By combining measurements of the proper motion of jet knots and the radial velocity reported in \citet{Cohen87}, \citet{Burrows96} estimated the jet inclination to be $90^\circ\pm3^\circ$. \citet{Pety06} and \citet{Louvet18} analyzed the inclination angle of the CO outflow axis, and it is $91^\circ\pm1^\circ$.

One possibility to explain the mismatch is to invoke a misaligned or warped structure in the innermost region of the disk, which would break the mirror symmetry in radiative transfer processes with respect to the disk midplane. In other words, if this is the case, the flux ratio at optical/near-IR is no longer a reliable tracer of the disk inclination anymore. 
\citet{Nealon19} found that an inner disk warp induced by a planet with a non-coplanar orbit can cast an appreciable shadow, affecting the flux ratio between the two surfaces.
They demonstrated that, even if the disk is exactly edge-on ($i=90^\circ$), the flux ratio can deviate significantly from unity in the presence of a warp.   
\citet{Villenave24} highlighted the flux inversion in IRAS 04302 that they interpreted using a misaligned inner disk. 
They showed that such a system can show the reversal of the flux ratio observed in IRAS 04302, which follows a very similar curve to HH 30 (see Figure \ref{fig:dwidth}). HH 30 is suggested to have a misaligned inner disk or a warped structure to explain the wiggling of the optical jet \citep{Anglada07, Estalella12} and the CO outflow \citep{Louvet18}, the above scenario would be a reasonable candidate to reconcile the issue. However, \citet{Louvet18} suggested a misalignment of $\sim1^\circ$ to explain the observed wiggling, and it is still unclear if such a small misalignment ($\sim1^\circ$) can fully reconcile an inclination mismatch greater than $1^\circ$. 

\subsection{Time variability of the reflection nebulae}

The disk reflection nebulae of HH 30 show variable asymmetry at optical/near-IR presumably due to variable asymmetric illumination from the central region of the system \citep{Stapelfeldt99}. \citet{Wood98} proposed that variable asymmetry is due to stellar hot spots created by magnetospheric accretion. Since the variable asymmetry mainly appears on the top surfaces, the hot spots are likely to exist only at the top of the stellar hemisphere \citep{Stapelfeldt99, Watson07}. 
Recent MHD simulations in the vicinity of the star also suggest that an asymmetry in the accretion flow between the top and bottom stellar hemispheres can arise naturally \citep{Takasao22}. The hot-spot model predicts a variability to occur with a timescale comparable to the stellar rotation period, which is around a few days. Photometric observations with the K2 mission revealed that HH 30’s light curve exhibits a quasi periodic behaviour with a period of 7.58 days \citep{Cody22}, in agreement with the earlier ground-based results by \citet{Duran2009}. 

The hotspot model predicts the color of the asymmetry to be blue due to its high temperature, yet the color remains neutral between V and I bands \citep{Watson07}. In contrast, despite the variable nature of HH 30, we do not find evidence of time variability either in their fluxes or lateral/vertical intensity profiles in the mid-IR between January 2023 and September 2023. Given the coarse sampling of the two epochs, we cannot rule out the presence of variability in the mid-IR. However, our result seems to be in line with the prediction of the hot-spot model in which we do not anticipate any variability in the mid-IR, as hot spots no longer contribute to mid-IR emissions as long as a thermal response of the disk due to a hot spot is negligible. 

\subsection{Origin of the spiral-like structure}

Since spiral structures are usually found more often in bright disks and around early type stars \citep{Garufi18,Benisty23}, it is interesting that we found a spiral-like feature in the disk around HH 30, an M0$\pm2$ star. In the following, assuming it is a spiral, we discuss potential mechanisms that may be responsible for driving it: (i) the inner binary, (ii) companion/flyby, (iii) an environment effect, (iv) WInDI (Warp-Induced Dust Instability), and (v) temperature-induced spiral. Although, given the amount of observational information, it is difficult to narrow down the mechanism, we discuss the pros and cons of each scenario.

{\it (i) Inner binary:} Since HH 30 has been suspected of being a binary \citep{Anglada07, Estalella12}, the binary-disk interaction might result in triggering a spiral in the circumbinary disk. To explain the wiggling of the optical jet, two binary configurations have been proposed, but in either case, the binary separation is no greater than 18 au \citep{Anglada07, Estalella12}. It is questionable if such a relatively compact binary can produce a spiral that is visible up to $\sim 300$ au seen in our JWST image. For example, the disk around HD 142527 is known to host a large-scale spiral extending up to 300 au \citep{Fukagawa06, Avenhaus14, Hunziker21}. \citet{Price18} found that the spiral could be naturally explained by gravitational interaction between the disk and the inner binary when a binary separation of 38.9 au is used. However, the binary separation of HD 142527 has been recently revised down to $\sim11$ au based on new orbital constraints with VLTI/GRAVITY, making it challenging to explain the spiral via the binary interaction \citep{Nowak24}. 

{\it (ii) Companion/Flyby:} 
The presence of a perturber orbiting either the interior or exterior of the spiral is another possibility \citep{Dong15, Dong16}.
\citet{Dong16} performed a comprehensive parameter survey of the viewing angle of the system with a perturber using 3D hydrodynamics and radiative transfer simulations.
Interestingly, some of their simulated images with a substellar-mass external perturber resemble the observed image of the HH 30 disk. 
In their simulations, the stellar companion is located at a distance twice larger than the disk's outer radius. However, we do not find a stellar companion in such a close proximity to the HH 30 region. 
Another possibility is a stellar flyby scenario \citep[see][for a review]{Cuello23}. 
In the case of HH 30, it is interesting to notice that there is a point source located in the north direction, and it is visible in both images taken in 1998 with HST and in 2023 with JWST, although it is not cataloged in GAIA DR3 \citep{Gaia16, Gaia23}.
In the JWST image, the projected angular separation between HH 30 and the source is $\approx 13\farcs6\approx 2000$ au by assuming a distance of 146.4 pc. We find that the separation is wider by $ 0\farcs30$ than the value measured in the 1998 image. The change in the angular separation translates into a velocity of $0\farcs012$/yr ($\sim8$ km s$^{-1}$), indicating the estimated perihelion passage was $\approx 1000$ yr ago if the source is a flyby source running away. This timescale is comparable to the timescale for which the spiral structure could be observed due to flyby \citep{Cuello20, Cuello23} and also smack on the value found  for the flyby of UX Tau A  by UX Tau C \citep{Menard20}.
However, we cannot exclude the possibility that the northern source is a background star. The HH 30 disk moves about 12 mas yr$^{-1}$ south and 8 mas yr$^{-1}$ east relative to the northern source, with an uncertainty of about 2 mas yr$^{-1}$ estimated based on $\approx1$ pixel uncertainty in the disk position. These values coincide with the direction of the proper motion of the L1551 cluster members whose median value is 18.943 mas yr$^{-1}$ south and 12.118 mas yr$^{-1}$ east \citep{Galli19}.

{\it (iii) Environment:} The spiral could also be triggered in a disk interacting with surrounding material/environment. Indeed, stars and disks are usually born in clusters \citep{Bate18}. \citet{Winter24} recently argued, from calculations, that 20--70\% of disks experience a replenishment of material from the surrounding environment. Observationnally, a late-infall of material on the disk has been proposed to explain some of the spiral structure seen in scattered light images \citep{Dullemond19, Kuffmeier21}. HH 30 is being associated with somewhat more prominent large scale nebula \citep{Burrows96}, and an infall event might have occurred in the past. In this case, the tail-like structure seen in Figure \ref{fig:nircam} might be interpreted as the remnant of such an event.
Note that there may be a CO emission counterpart to the tail-like structure, but its kinematics remains inconclusive because the velocity is close to the systemic velocity \citep{Lopez24}.
The disks interacting with environments however often show an intricate structure, such as a multi-spiral/flocculent pattern, as in AB Aur \citep{Hashimoto11, Boccaletti20} or HD 34700 \citep{Monnier19, Uyama20, Columba24, Minghan24} or a rather intricate scattered light morphology, as in WW Cha \citep{Garufi20} or SU Aur \citep{Ginski21}, whereas the HH 30 disk has a relatively ordered structure. Thus further studies are needed to see if the observed HH 30's tail could still be attributed to such an interacting event. 
It is worth pointing out that AB Aur, HD 34700, and SU Aur are all high-luminosity stars that illuminate their surroundings to much larger distances than HH 30 does, and hence, the comparison between those disks and HH 30 has to be interpreted with some care.

{\it (iv) WInDI:} Recently, \citet{Aly24} found a new instability that occurs in a warped disk. The instability is called WInDI (Warp-Induced Dust Instability). WInDI causes a broken ring in the dust distribution that resembles a spiral structure. Since the HH 30 disk has been suggested to have a warped or misaligned structure in the innermost region based on the wiggling of the optical jet and CO outflows \citep{Anglada07, Estalella12, Louvet18}, WInDI may therefore be an interesting 
candidate to explain the observed spiral-like structure. However, the spiral seen in HH 30 is visible at a rather high altitude in the disk, e.g., $z/r\sim0.5$ at $r\sim1\arcsec\approx147$ au, and it is unclear whether WInDI can imprint such a spiral signature this high above the disk midplane. Further investigation is needed. 

{\it (v) Temperature-induced spiral:} \citet{Montesinos16} showed using 2D hydrodynamics simulations that the local temperature changes due to shadowing lead to the formation of spiral arms in a self-gravitating disk \citep[see also][]{Su24, Zhang24, Ziampras24}. 
Although the spiral driven by this mechanism would be too weak to efficiently trap pebbles, \citet{Montesinos16} and \citet{Su24} suggested the possibility that the structure could be detectable in scattered light images. If this mechanism is at work, the spiral-like structure seen in HH 30 would be expected to rotate with the pattern speed of the shadow and might explain the variable asymmetry \citep{Stapelfeldt99, Watson07}.

Although some of the mechanisms might explain the observed feature, such as a late infall or a flyby scenario, it is not immediately clear which scenario is most likely given the limited observational information available. Furthermore, the spiral nature of the feature should be interpreted with some care for the moment because it may instead be related to another type of substructure, such as a ring or a shadow. A detailed analysis from both the observation and modeling sides is necessary to draw more robust conclusions. 

\subsection{Absence of an X-shaped mid-IR feature in HH 30}

Our MIRI images of HH 30 do not reveal the presence of an extended mid-IR emission component, such as the X-shaped emission seen in Tau042021 \citep{Duchene24}. We remind readers that the X-shaped emissions seen in our 7.7-\micron~MIRI images are the cruciform artifact of the MIRI imager \citep{Gaspar21} (see section \ref{sec:ref}) and not a real structure. In that case, the two branches of the feature are orthogonal to each other.

\citet{Duchene24} found an X-shaped emission component in the MIRI images of Tau 042021 that exists well above the scattered light surfaces of the disk, which is presumably connected to disk winds.
A semi-opening angle of the X-feature is measured to be $\sim55^\circ$, so clearly not the cruciform artifact. A similar X-shaped feature has also been reported in the MIRI IFU data for the same source \citep{Arulanantham24}, but with a significantly different semi-opening angle of $35^\circ\pm 5^\circ$. \citet{Sturm23c} and \citet{Sturm24} found a similar "X"-shaped feature in HH 48 NE, and it is shown to be associated with H$_2$ emission. Both Tau 042021 and HH 48 show a collimated jet to which those X features may be correlated.
However, despite the fact that HH 30 hosts a strong collimated jet \citep{Anglada07, Estalella12, Ai24} and an outflow \citep{Pety06, Louvet18, Lopez24}, we do not find these mid-IR features in the images. Although the number of samples is still limited, the current observations so far seem to indicate that the occurrence of the X-shaped emission feature does not correlate with the collimated jet.

Since the origin of the X-shaped feature seen in some edge-on disks still remains unclear, the absence of it in the HH 30 disk would serve as an interesting test case to verify the mechanisms for producing those features.

\section{Conclusions} \label{sec:conclusion}

We have presented new broadband imaging observations of the HH 30 disk with JWST/NIRCam and MIRI together with the new ALMA Band 6 obtained at maximum resolution. By combining those datasets with HST images at optical/near-IR, we study the multiwavelength disk appearance of the system. We carried out radiative transfer simulations and the results were then compared with the observations. The primary findings of this paper are as follows. 

\begin{itemize}

\item With JWST/NIRcam and MIRI, we detected the disk reflection nebulae bisected by the disk midplane from 2~\micron~to 7.7~\micron~(or 12.8~\micron). The NIRCam/F200W (2 \micron) image shows a number of structures, such as a jet, a conical structure, a tail, and a spiral-like structure, some of which are also visible at other filters. The collimated jet is particularly bright in the MIRI/F1280W filter (12.8\micron) with an emission knot detected. 

\item Taking advantage of a broad wavelength coverage, extended by adding HST archival images, we have shown that the separation between the reflection nebulae varies significantly with wavelength. The nebular separation is greatly reduced from $\sim0\farcs9$ at 0.8~\micron~to $\sim0\farcs5$ at 2.0~\micron, but it remains nearly the same at longer wavelengths.

\item We found that the flux ratio of the bottom and top surfaces of the disk remains nearly constant for $\lambda\le2~\micron$; however, it changes significantly at $\lambda>2~\micron$. At 2~\micron, the flux ratio is $\sim0.3$, while at 12.8~\micron~it is $\sim1$. 
Current simple models based on a smooth, flared disk morphology cannot explain such a peculiar wavelength dependence. 

\item Efficient mid-IR light scattering by dust grains is mandatory to reproduce the observed bright mid-IR emission, requiring micron-sized grains. Furthermore, to account for the observed geometrically thick appearance of the mid-IR reflection nebulae, these large grains must be abundant in the surface region. By running radiative transfer models, we showed that grains of about 3 \micron~in radius or larger must be fully vertically mixed in the outer disk surface at about 50--100 au from the central star. 

\item With the MIRI/F1280W filter, we detected a mid-IR jet of HH 30, likely tracing [NeII],  extending up to $\sim8\arcsec$ ($\approx1000$ au above the midplane). Its counter-jet component is absent in the MIRI image. In addition, we detected for the first time a proper motion of a knot seen in the mid-IR jet of HH 30. The knot moves with a velocity of $0\farcs174$/yr, translating into 121 km s$^{-1}$. This is similar to the velocity of knots seen in the optical jet \citep{Estalella12}.

\item We detected a conical structure bracketing the collimated jet in the NIRCam/2-\micron~image. It has a semi-opening angle measured from the jet axis of $\sim14^\circ$. Assuming that the trajectory of the outflow is characterized by a linear function down to the midplane, we estimated the launching radius of the conical outflow to be less than $\sim16$ au.

\item With high spatial resolution ALMA Band 6 observations, 
we have spatially resolved the dust continuum emission of the HH 30 disk. The emission shows a flat and boxy morphology, which is characterized by a uniform brightness along the major axis and a bell-shaped profile along the minor axis. The uniform brightness suggests that the disk is optically thick at the ALMA Band 6. The full thickness and diameter of the disk emission is about $0\farcs22$ and $1\farcs6$, respectively, although the disk thickness does not represent the scale height.

\item The model with a scale height of mm grains of 4.7 au at 100 au from the central star can reproduce the vertical thickness of the 1.3-mm image. Even if the disk mass is increased up to an uncomfortably large mass, the scale height of mm grains still seems to point to greater than 1.5 au at 100 au. Therefore, the pebbles in the HH 30 disk do not settle as efficiently as those in HL Tau or Oph 163131. 

\item The uniform brightness of the dust continuum emission at the ALMA Band 6 suggests that the disk is nearly exactly edge on $i\gtrsim89^\circ$. In contrast, the inclination angle derived from the bottom/top flux ratio at optical/near-IR is $84^\circ$--$86^\circ$. The flux ratio is time-varying \citep{Watson07}; however, its effect cannot fully reconcile the discrepancy. Although the inclination angle estimated by the flux ratio could be affected by some non-uniform illumination effects, further study is needed to clarify whether such an effect can fully reconcile the discrepancy. 

\item Despite the variable optical/near-IR scattered light of the HH 30 disk \citep{Stapelfeldt99, Watson07}, we do not find evidence of time variability of a photometric flux and intensity profiles at mid-IR wavelengths taken in January 2023 and September 2023 except for the movement of the emission knot associated with the jet seen in F1280W filter. However, it is still possible that the lack of variability is due to obviously insufficient epoch sampling.
\end{itemize}

Our observations and simple radiative transfer modeling have raised new questions about HH 30 that will need to be answered by the future observations and detailed modeling: (i) the origin of grains of a few microns in size in the outer disk surface regions revealed by our NIRCam and MIRI observations, (ii) the origin of the changing top-to-bottom flux ratio of the bi-reflection nebulae from optical to mid-IR as well as an apparent conflict of the disk inclination angle inferred from optical/near-IR and millimeter-wave observations (iii) the driving mechanism of the nested conical outflow \citep[e.g.,][]{Pascucci24}, (iv) the absence of the counterjet in the 12.8 $\mu$m filter, and (v) the absence of spatially extended mid-IR component unlike some other edge-on disks (Tau 042021, Oph 163131) \citep{Duchene24, Villenave_Oph163131}, (vi) the absence of mid-IR time variability between our two epochs of MIRI observations and its relation to optical variability \citep{Cody22} and variable scattered light \citep{Stapelfeldt99, Watson07} of the system.

\acknowledgments

We thank Rebecca Nealon, Nicol\'as Cuello, Ruobing Dong, and Shinsuke Takasao for several useful discussions.
This work benefitted from funding from the European Research Council (ERC) under the European Union's Horizon Europe research and innovation program (grant agreement No. 101053020, project Dust2Planets, PI F. M\'enard). MV acknowledges funding from the European Research Council (ERC) under the European Union's Horizon Europe research and innovation program (grant agreement No. 101039651, project DiscEvol, PI G. Rosotti).
MV, KRS., GD, and SGW. acknowledge funding support from JWST GO program \#2562 provided by NASA through a grant from the Space Telescope Science Institute, which is operated by the Association of Universities for Research in Astronomy, Incorporated, under NASA contract NAS5-26555.
A.R. has been supported by the UK Science and Technology Facilities Council (STFC) via the consolidated grant ST/W000997/1 and by the European Union’s Horizon 2020 research and innovation programme under the Marie Sklodowska-Curie grant agreement No. 823823 (RISE DUSTBUSTERS project).
This paper makes use of the following ALMA data: ADS/JAO.ALMA\#2017.1.01701.S ALMA is a partnership of ESO (representing its member states), NSF (USA) and NINS (Japan), together with NRC (Canada), MOST and ASIAA (Taiwan), and KASI (Republic of Korea), in cooperation with the Republic of Chile. The Joint ALMA Observatory is operated by ESO, AUI/NRAO and NAOJ. The National Radio Astronomy Observatory is a facility of the National Science Foundation operated under cooperative agreement by Associated Universities, Inc.
This work has made use of data from the European Space Agency (ESA) mission
{\it Gaia} (\url{https://www.cosmos.esa.int/gaia}), processed by the {\it Gaia}
Data Processing and Analysis Consortium (DPAC,
\url{https://www.cosmos.esa.int/web/gaia/dpac/consortium}). Funding for the DPAC
has been provided by national institutions, in particular the institutions
participating in the {\it Gaia} Multilateral Agreement. The specific observations analyzed can be accessed via \dataset[doi: 10.17909/rrq0-qx18]{https://doi.org10.17909/rrq0-qx18} for the HST data and \dataset[doi: 10.17909/7m4d-vz55]{https://doi.org/10.17909/7m4d-vz55} for the JWST data.

\software{\texttt{numpy} \citep{harris20}, \texttt{matplotlib} \citep{hunter07}, \texttt{RADMC-3D v2.0} \citep{Dullemond12}, \texttt{OpTool} \citep{Dominik21}, \texttt{WebbPSF} \citep{Perrin14}}

\newpage

\appendix

\section{Spike subtraction} \label{sec:spikesub}

\begin{figure*}[t]
\begin{center}
\includegraphics[width=0.49\linewidth]{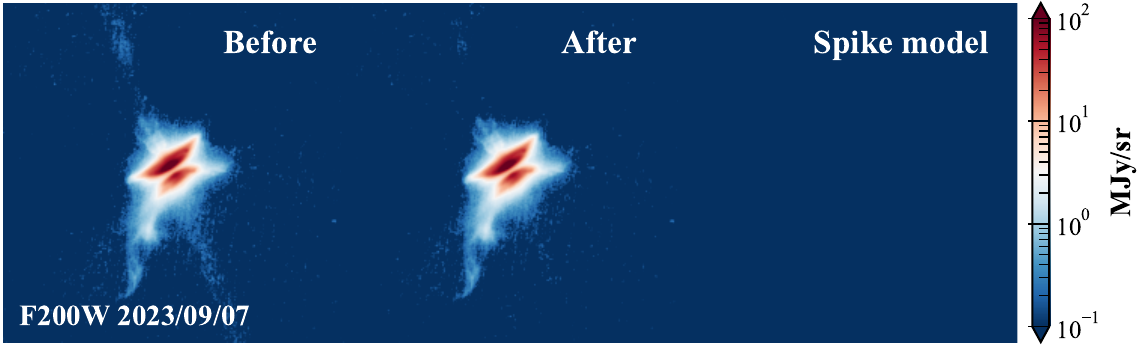}
\includegraphics[width=0.49\linewidth]{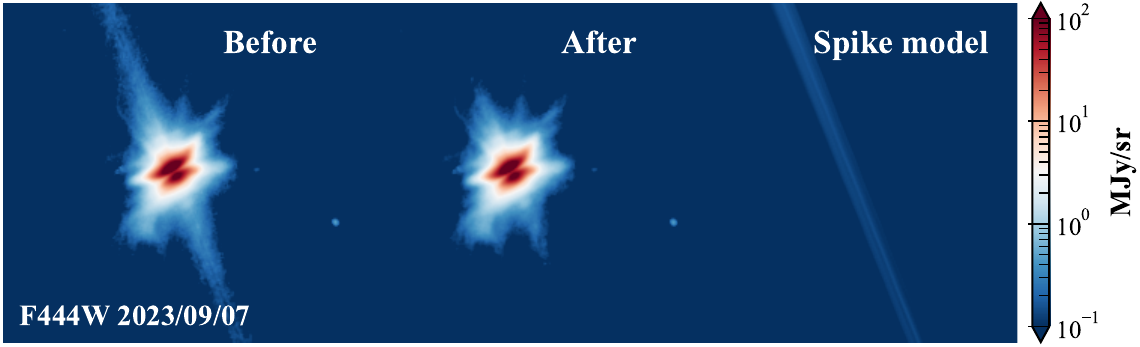}
\includegraphics[width=0.49\linewidth]{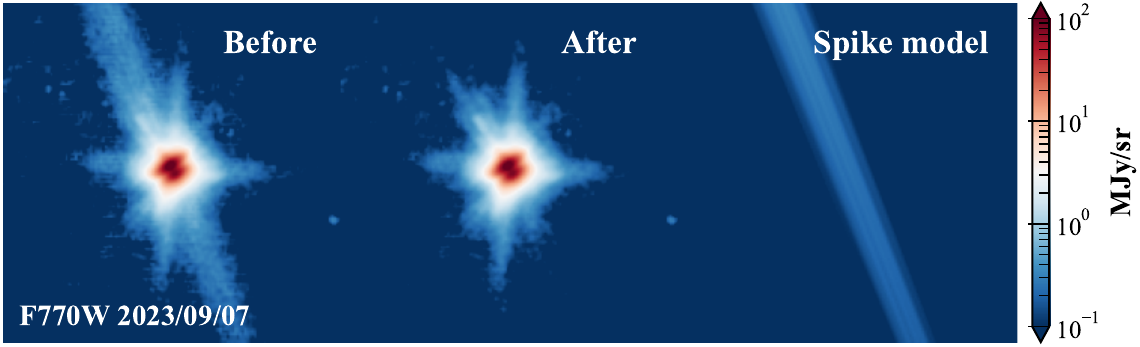}
\includegraphics[width=0.49\linewidth]{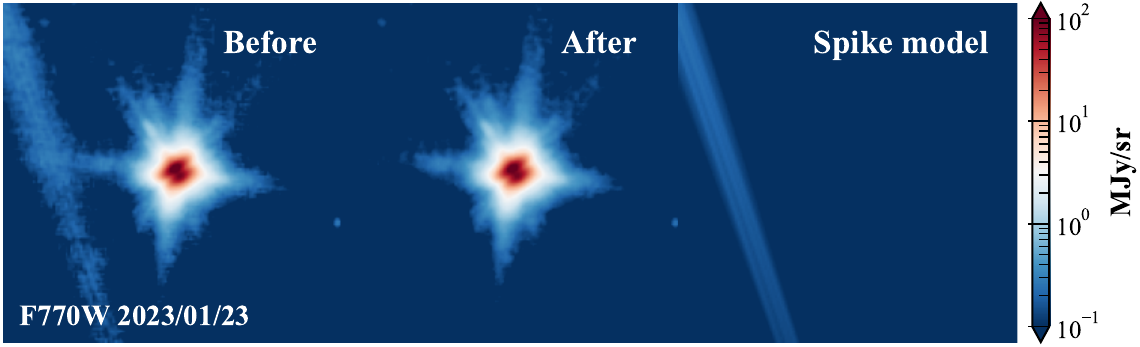}
\includegraphics[width=0.49\linewidth]{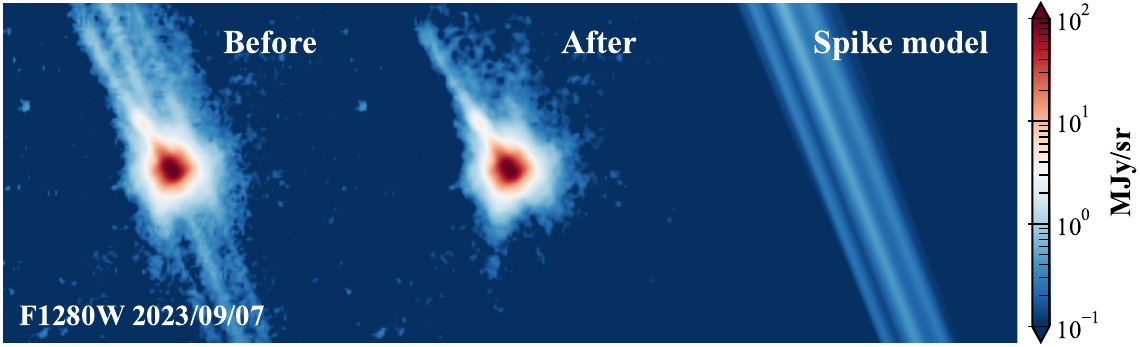}
\includegraphics[width=0.49\linewidth]{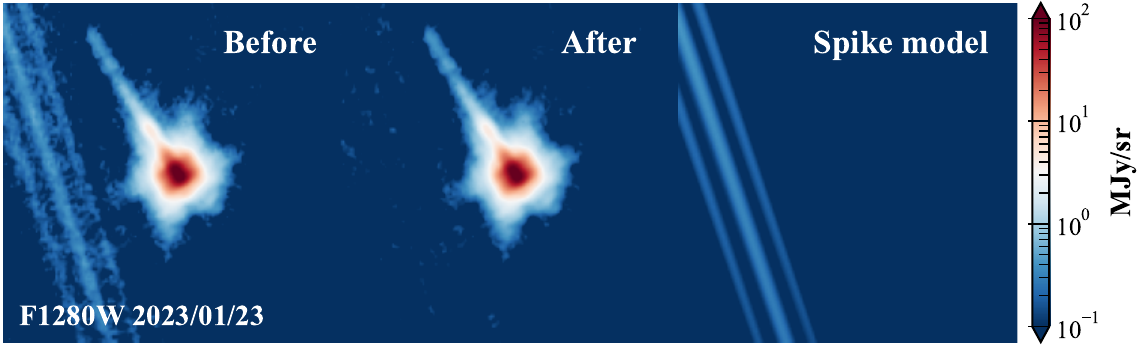}
\includegraphics[width=0.49\linewidth]{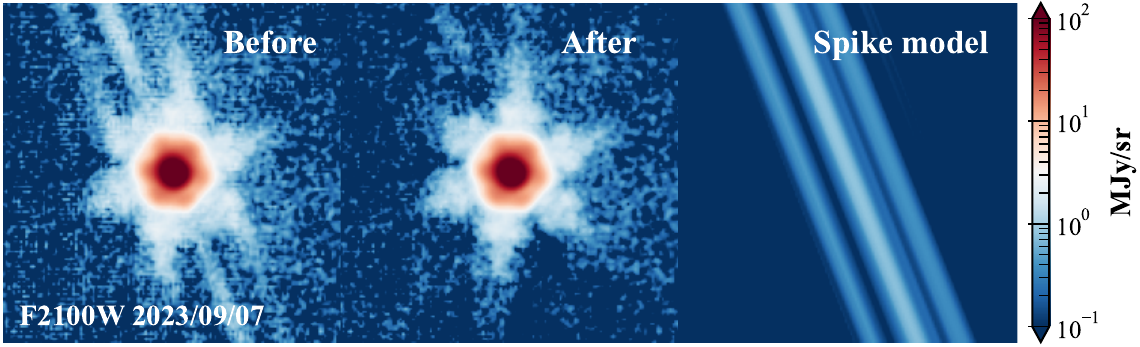}
\includegraphics[width=0.49\linewidth]{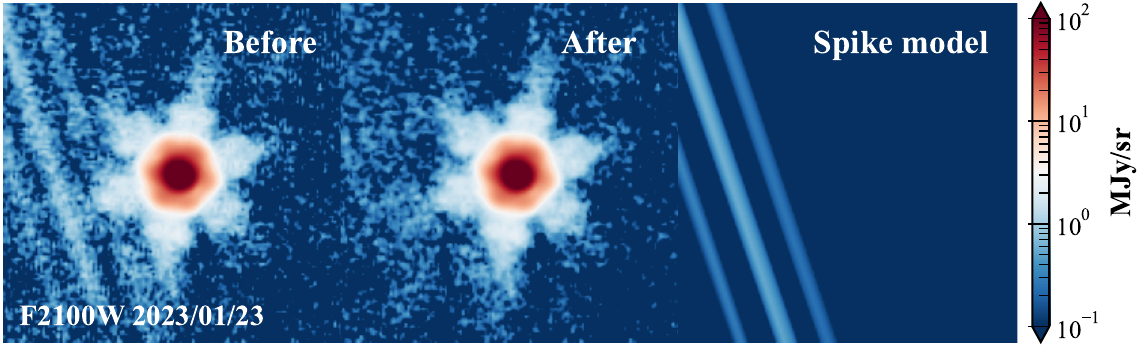}
\caption{Subtraction of a diffraction spike in the HH 30 region. Each panel consists of three panels: from left to right, before and after the diffraction spike subtraction and the diffraction spike model used in the subtraction. Each panel has the field of view of 15\arcsec$\times$15\arcsec. The images are shown with north-up and east-left convention.}
\label{fig:spikesub}
\end{center}
\end{figure*}

We outline procedures used to subtract the diffraction spike from XZ Tau that comes on top of the HH 30 disk (Figure \ref{fig:spikesub}). 
We first rotated each image so that the diffraction spike lies horizontally and extracted the pixel values along the spike on a row-by-row basis. 
We applied masking to the bright point sources detected with the \texttt{photoutils} package and also for all pixels within a rectangular region with the size of 14\arcsec ($\sim2050$ au)~centered on HH 30 to avoid fitting the disk signal itself (e.g., jet, outflow). 
The masked image was used to extract the spike intensity profile along it, and it was then fitted by a quadratic function either in a logarithmic space if all the pixel values are positive or in a linear space if they contain both positive and negative values. 
The fitting function is then used to subtract the diffraction spike from the observed image. By performing the above analysis for all rows containing the diffraction spike, we obtain a spike-subtracted image. 

The above issue does not happen for MIRI datasets obtained in January 2023, but one of the diffraction spikes appears close to HH 30. Therefore, we applied the same reduction procedure not only to the September dataset but also to the January dataset. Figure \ref{fig:spikesub} shows the comparison of before and after the spike subtraction around the HH 30 region as well as the spike model we created. 

\section{All JWST images} \label{sec:jwstall}

\begin{figure*}[tbp]
\centering
\includegraphics[width=\linewidth]{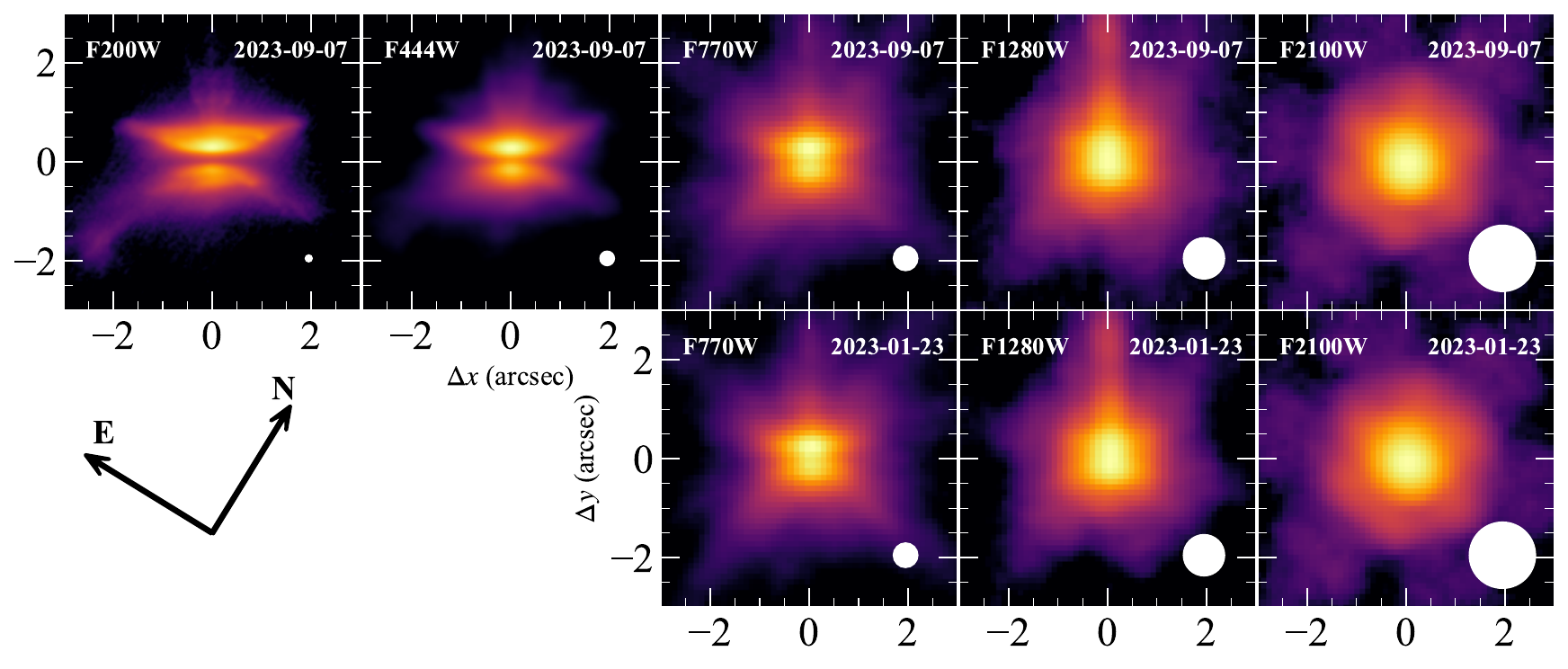}
\caption{All JWST images obtained in our program with observational wavelengths from 2~\micron~to 21~\micron. All images are shown with the same angular scale and a logarithmic stretch. The corresponding PSF size is shown in the bottom right of each plot.}
\label{fig:jwst_all}
\end{figure*}

\begin{figure}[tbp]
\centering
\includegraphics[width=0.5\linewidth]{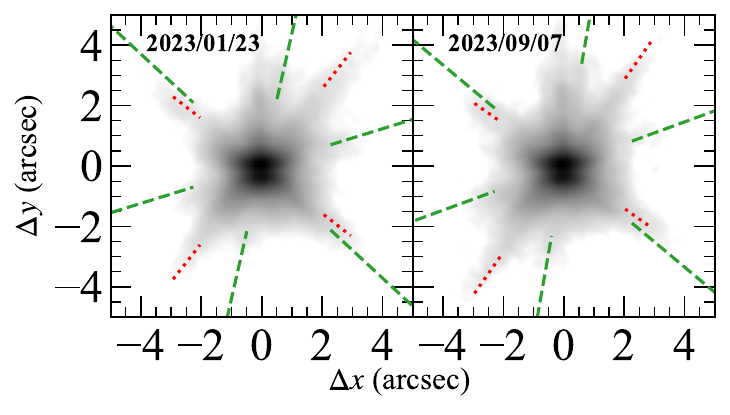}
\caption{MIRI 7.7~\micron~images and orientations of possible artifacts. Two images are centered on the brightest pixel. Dotted and dashed lines represent the orientation of the cruciform artifact \citep{Gaspar21} and the diffraction spikes. The alignment of the native detector orientation and the diagonal feature indicates its artifact nature. The red dotted and green dashed lines represent the orientations of the detector and telescope's diffraction spikes to which direction an excess emission due to artifacts could appear}.
\label{fig:artifacts}
\end{figure}

Figure \ref{fig:jwst_all} shows a gallery of all JWST images obtained in our program.
In the MIRI 7.7~\micron~images, diagonal features show up, illustrated more clearly in Figure \ref{fig:artifacts}. 
However, these diagonal features are unlikely to be real because their orientation aligns well with the detector orientation (Figure \ref{fig:artifacts}). It seems to be due to a detector artifact, known as the cruciform artifact \citep{Gaspar21}. The cruciform artifact is one of the known artifacts of the MIRI instrument, and it is a consequence of the internal scattering of incoming light within the MIRI detector, resulting in creating a cruciform pattern for $\lambda\le10~$\micron.
This is distinct from the X-feature reported in \citet{Duchene24} for Tau 042021, which is not aligned with known instrument artifacts and is believed to result from the emission of H$_2$ and/or PAH \citep{Arulanantham24}.

\section{Dust opacity law} \label{sec:opac}
\begin{figure*}[t]
\begin{center}
\includegraphics[width=\linewidth]{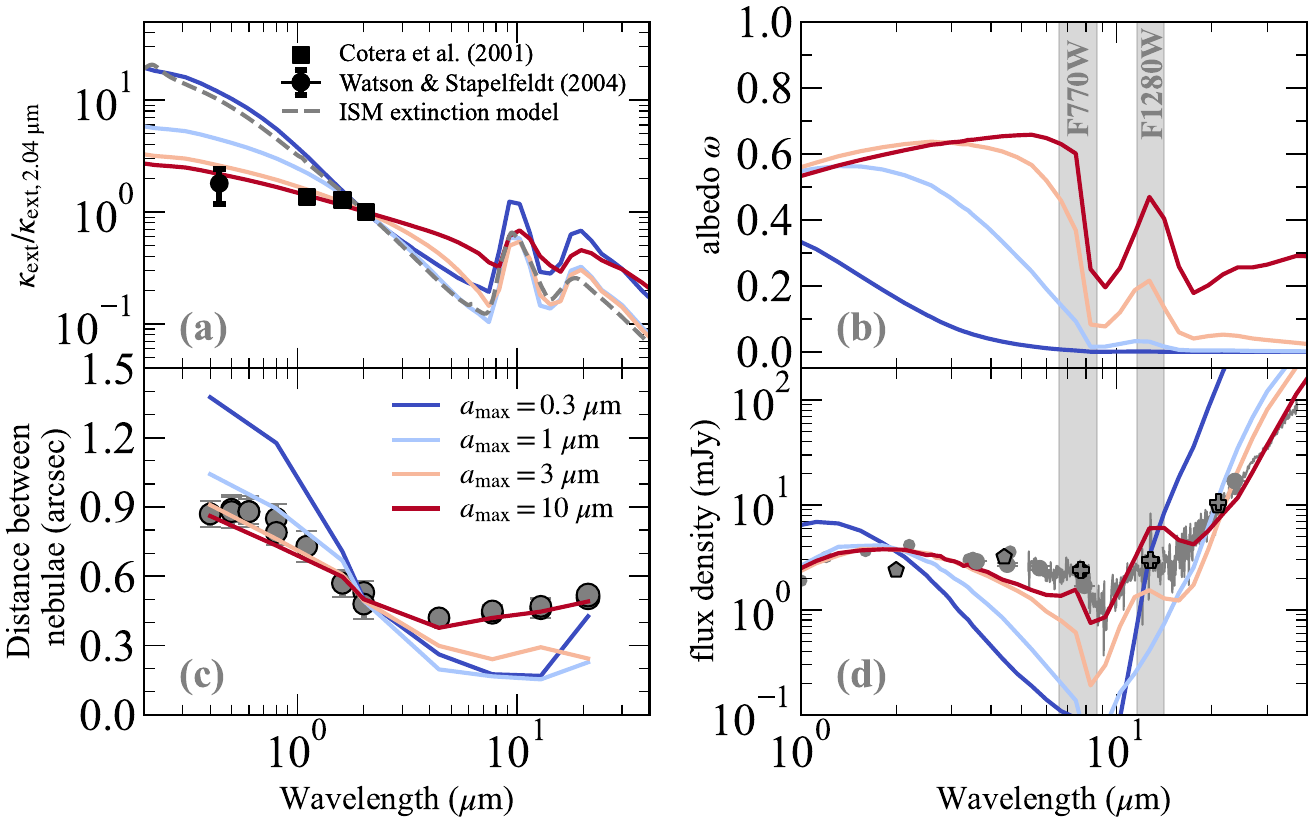}
\caption{
(a) The extinction opacity laws for $\amax=0.3$, 1, 3, 10~\micron~normalized to 2.04~\micron. 
The gray dashed line represents the extinction curve of the interstellar grains \citep[with $R_\mathrm{V}=3.1$, see][for more details]{Draine03}.
The square and the circle symbols represent the opacity law of dust grains in the surface layer of the HH 30 disk estimated in \citet{Cotera01} and \citet{Watson04}, respectively.
(b) The single scattering albedo of the corresponding models.
(c and d) The nebular separation and SED for the opacity models that are shown in the panels (a) and (b).}
\label{fig:opac}
\end{center}
\end{figure*}

\citet{Cotera01} and \citet{Watson04} studied the dust opacity law for grains in the HH 30 disk surface based on the nebular separation with earlier HST images. In this section, we demonstrate that our model is consistent with the previous constraints of the opacity law.

To facilitate a more direct comparison between the opacity model and the nebular separation, we performed radiative transfer simulations with a simplified setup. In contrast to simulations presented in Section \ref{sec:irmodel}, where the dust opacity varies depending on disk radius and vertical height due to dust settling, here we consider a fully mixed model, where the opacity model is the same everywhere within the disk. 
Figures \ref{fig:opac}(a, b) show the extinction opacity law and the single scattering albedo for a grain-size distribution with the maximum grain radii of $\amax$=0.3, 1, 3, or 10~\micron. The resultant nebular separation and SED corresponding to each opacity model are shown in Figure \ref{fig:opac} (c,d). 

As the maximum grain radius is increased, the extinction opacity law becomes flatter (a), reducing and increasing the nebular separation at $\lambda\le2~\micron$ and $\ge2~\micron$ (c). Consequently, the nebular separation becomes less chromatic. The observed nebular chromaticity is well reproduced by an opacity model with $\amax=10~\micron$.
It might be interesting to compare how those extinction opacity laws are compatible with a typical interstellar extinction curve \citep{Draine03}. 
The extinction curve of $\amax$=0.3~\micron~ is similar to that of ISM grains, whereas that of $\amax=10~\micron$ is much flatter. Therefore, despite the fact that we probe the very surface region of the disk, grains are much larger than the interstellar grains. 
Therefore, our opacity model is fully consistent with the opacity law found in earlier studies based on HST images of HH 30 \citep{Cotera01, Watson04}.

\bibliography{cite}

\end{document}